\documentclass[letterpaper]{JHEP3}
\pdfoutput=1

\usepackage{amsmath}
\usepackage{amssymb}
\usepackage{amsfonts}
\usepackage{lscape, graphicx}
\usepackage{cite}       
\usepackage{bm}         
\usepackage{verbatim}

\usepackage[all]{xy}
\advance\parskip 1.0pt plus 1.0pt minus 2.0pt   
\def\href#1#2{#2}	

\def\N{{\cal N}}

\def\R{{\mathbb R}}
\def\S{{\mathbb S}}

\def\tr{{\rm tr}}

\def\Z{{\mathbb Z}}

\def\Dslash{{\rlap{\raise 1pt \hbox{$\>/$}}D}}

\newcommand{\beq}{\begin{equation}}
\newcommand{\eeq}{\end{equation}}
\newcommand{\beqa}{\begin{eqnarray}}
\newcommand{\eeqa}{\end{eqnarray}}

\def\ltap{\ \raise.3ex\hbox{$<$\kern-.75em\lower1ex\hbox{$\sim$}}\ }
\def\gtap{\ \raise.3ex\hbox{$>$\kern-.75em\lower1ex\hbox{$\sim$}}\ }
\def\gl{\ \raise.5ex\hbox{$>$}\kern-.8em\lower.5ex\hbox{$<$}\ }
\def\roughly#1{\raise.3ex\hbox{$#1$\kern-.75em\lower1ex\hbox{$\sim$}}}

\title{ Microscopic Structure of Magnetic Bions }

\author
    {
    {
    \def\href#1#2{#2}	
    Mohamed M. Anber\footnote{\email{manber@physics.utoronto.ca}}~
    and Erich Poppitz\footnote{\email{poppitz@physics.utoronto.ca}}~
           \\ Department of Physics, University of Toronto,
    Toronto, ON M5S 1A7, Canada
            }
    }%
    
    \abstract{
    
    \smallskip
    
    \smallskip
    
    {\small{

Magnetic bions---stable bound states of monopoles and twisted  (``Kaluza-Klein")  monopoles, carrying two units of magnetic charge---have been shown to be the leading cause of confinement and mass gap
in four-dimensional  gauge theories with massless adjoint fermions compactified on $\R^{1,2} \times \S^1$, at least at small $\S^1$. In this paper, we study in detail the bion mechanism and the  scales involved for an $SU(2)$ gauge group, using  traditional QCD instanton methods.  We represent the vacuum functional as the partition function of a bion-anti-bion plasma and obtain the next-to-leading dependence of the mass gap  on the $\S^1$ size $L$ at fixed strong-coupling scale $\Lambda$. We find that, at  small $\Lambda L$, the mass gap is  an increasing function of $L$  for theories with four massless adjoint Weyl  fermions, a case left undetermined by the previous leading-order analysis, and comment on the approach to $\R^4$.}}}

\begin{document}

\maketitle

\section{Introduction and outline}

It was recently realized that compactifying general four-dimensional gauge theories on $\R^{1,2} \times \S^1$, where $\S^1$ is a non-thermal circle of size $L$, offers the opportunity to study many difficult questions of nonperturbative gauge dynamics, such as confinement and the generation of mass gap, in a theoretically controlled setting \cite{Unsal:2007vu,Unsal:2007jx}. The first studies of gauge dynamics in non-thermal circle compactifications were performed in supersymmetric theories and yielded many interesting 
results  \cite{Seiberg:1996nz,Aharony:1997bx,Davies:1999uw}.\footnote{Even in SUSY, it appears to us that circle compactifications have not been fully utilized yet, see, e.g.,  \cite{Poppitz:2009kz}.} More recently, it was shown that non-thermal circle compactifications offer a calculable window into the dynamics of general nonsupersymmetric  gauge theories as well\cite{Unsal:2006pj,Shifman:2008ja,Unsal:2008ch}. 

For general gauge theories, obtaining theoretical control at small $L$ requires the use of  particular double-trace deformations \cite{Shifman:2008ja}. 
However, in the most ``friendly" examples calculability at small $L$ is ``automatic." This is the case in  four-dimensional gauge theories with multiple massless adjoint Weyl fermions \cite{Unsal:2006pj} compactified on $\R^{1,2} \times \S^1$, the subject of this 
paper.\footnote{Small-$L$ calculability is automatic  also  in some more ``exotic" gauge theories, see \cite{Poppitz:2009tw}.} 

\" Unsal showed \cite{Unsal:2007vu,Unsal:2007jx}  that theories with $n_f$ massless Weyl adjoints 
generate a mass gap and exhibit confinement of electric charges via a generalization of the three-dimensional (3D) Polyakov mechanism \cite{Polyakov:1976fu} to a locally four-dimensional (4D) theory, i.e., at $L$ small but nonzero. The topological objects responsible for the screening of magnetic fields and for the formation of a confining electric flux tube  are not magnetic monopole-instantons, as in the Polyakov  model, but weakly bound states of monopole-instantons and twisted monopole-instantons (note that the latter do not exist in the 3D limit, hence this is a locally 4D mechanism). The binding into stable objects  of magnetic charge two, the 	``magnetic bions," occurs due to fermion-induced attraction between the constituent monopole-instantons. The realization of  this picture of confinement is an important result, as it shows that a well-known 3D mechanism of confinement extends into 4D, at least for sufficiently small $L$,  and that massless fermions do not switch off Polyakov's mechanism, as previously thought. This is also the first time confinement is  analytically demonstrated   in a non-supersymmetric, continuum,  locally 4D theory. 

{\flushleft{T}}he purpose in this paper is two-fold:
\begin{enumerate}
\item We want to study the binding and exhibit the scales relevant to the bion gas using methods traditional to QCD instanton calculations \cite{Schafer:1996wv}, where fermion-induced interactions between instantons play a role, e.g., in the instanton-liquid model, and we explain why the method used in \cite{Unsal:2007jx} also applies. For the non-experts, we also wish to review and collect the relevant classical solutions using a unified notation and terminology, as both vary wildly in the literature and can be sometimes confusing.

\item We want to study the behavior of the mass gap for gauge fluctuations upon increasing the $\S^1$ size $L$ at fixed strong-coupling scale $\Lambda$. The existing calculation \cite{Poppitz:2009uq}, which takes into account only the leading 't Hooft suppression factors, shows that for $n_f < 4$, the mass gap is an increasing function of $L$, while for $n_f > 4$, the mass gap decreases upon increasing $L$. The $n_f=4$ case was left undetermined by the leading-order calculation.\footnote{While this is tangential to the subject of  this paper, we note that the $n_f =4$ theories have received some attention due to interest in ``minimal walking technicolor" models of electroweak symmetry breaking, see \cite{Andersen:2011yj} and references therein. Note also that the $n_f=4$ theory is nothing but $\N =4$ supersymmetric Yang-Mills with the scalars decoupled, but the $SU(4)_R$ chiral symmetry left intact. } Fixing this indeterminacy requires understanding and calculating the relevant corrections to the leading ``bion-instanton" result.  
\end{enumerate}
 
{\flushleft{T}}his paper is organized as follows. 

\smallskip

In Section \ref{perturbative}, we define our notation and the class of $SU(2)$ gauge theories with $n_f$ massless adjoint Weyl fermions that we study. We discuss their behavior upon compactification on small $\R^{1,2}\times \S^1$, show that the theory with $n_f>1$ ``abelianizes," and present the effective lagrangian describing the perturbative dynamics at distances $\gg L$.

Section \ref{instantons} is a review, only useful to the non-experts, aiming to collect the various solutions (noting that sometimes the terminology used in the literature can be confusing).   In Section \ref{instantons2}, we discuss the classification of the  finite-action Euclidean  topological solutions in  theories on $\R^3 \times \S^1$.
  In Section \ref{monopoleinstantons}, we present the monopole-instanton solutions independent of the $\S^1$ coordinate. Their magnetic, electric, gauge, and ``Higgs" fields in the BPS limit are given in Section \ref{bpsmonopoles}. In Section \ref{kkmonopoles}, we explain the construction of the ``twisted" (or ``Kaluza-Klein") tower of monopole-instantons, give the BPS-limit magnetic, electric, gauge, and ``Higgs" field profiles of the lowest action member of the infinite family of twisted solutions, and describe the long-distance interactions between the various monopole-instanton solutions in the BPS limit. In Section \ref{calorons}, we  briefly review the relation of the monopole-instantons and twisted monopole-instantons to the BPST instanton solutions with nontrivial holonomy  (``calorons"). 

 Section \ref{dynamicsofconfinement} contains the main body of the paper. 
 In Section \ref{vacuumatsmalll}, we discuss the relevance of the various classical solutions reviewed in Section \ref{instantons} to confinement and the generation of mass gap, stress the relevance of magnetic bions, and review their quantum numbers.
 
In Section \ref{bionplasma}, we show how representing the partition function of the theory as that of a  classical bion plasma (assuming the existence of stable magnetic bions) leads to a ``Polyakov-like" magnetic bion mechanism of confinement. 

In Section \ref{bionstructure}, we  calculate the  forces between the constituents of a bion and give an expression for the single bion partition function. The magnetic interaction energy between the constituent monopole-instantons and the fermion-induced attraction are  calculated in Sections \ref{computationofsint} and \ref{computationofdet}, respectively.
These are collected in Section \ref{computationofzbion}, where the final expression for the bion partition function is given and the bion fugacity is calculated. The various scales relevant to the magnetic bions are also shown on Figure \ref{fig:bionsdraw}.

In Section \ref{dualphotonmass}, we study the dependence of the dual photon mass (``mass gap for gauge fluctuations") on the $\S^1$ size $L$ for various numbers of Weyl adjoints $n_f$. We discuss the relation to  previous small-$L$ ``estimates" of the conformal window in  theories with multiple adjoint fermions and the consistency with recent lattice results.

In the final Section \ref{discussion}, we summarize our results and discuss the likely behavior of the mass gap as $L \rightarrow \infty$.

\section{Perturbative treatment of QCD(adj) on $\mathbf{\R^3 \times \S^1}$}
\label{perturbative}

We consider $4D$ $SU(2)$ Yang-Mills theory with $n_f$ Weyl fermions in the adjoint representation. These QCD-like (vector) theories are denoted QCD(adj). The action for $SU(N)$ QCD(adj) defined on $\R^3\times \S^1$ is \cite{Unsal:2007vu,Unsal:2007jx}:
\begin{eqnarray}\label{basic lagrangian}
S=\int_{\R^3\times \S^1}\mbox{tr}\left[-\frac{1}{2g^2}F_{MN}F^{MN}+ 2 i\bar \lambda^I \bar\sigma^MD_M\lambda_I \right]\,,
\end{eqnarray}
where we use the signature $\eta_{\mu\nu}=\mbox{diag}(+,-,-,-)$, $F_{MN}$ is the field strength, $D_M$ is the covariant derivative, $I$ is the flavor index, $\lambda_I=\lambda_{I,a}t_a$, $a=1,...,N^2-1$,  $\sigma_M=(1,\tau_i)$, $\bar\sigma_M=(1,-\tau_i)$, $1$ and $\tau_a$ are respectively the identity and Pauli matrices, and the generators $t_a$ are normalized as $\mbox{Tr}\,t_at_b=\frac{\delta_{ab}}{2}$. The upper case Latin letters $M\,, N$ run over $0,1,2,3$, the lower case Latin letters $i\,,j$ run over $1,2,3$, and the Greek letters $\mu\,,\nu$ run over $0,1,2$. The components $0$ and $3$ denote time and the compact dimension, respectively. Thus $x^3 \equiv x^3 + L$, where $L$ is the circumference of the $\S^1$ circle.

 Notice that  we are compactifying a spatial direction and the time direction $\in \R^3$,  in other words $\R^3\equiv \R^{2,1}$. In the following, we restrict our attention to $N=2$, and  take $t_a=\tau_a/2$.   The quantum theory has a dynamical strong scale $\Lambda$ such that, to two-loop order, we have:
\begin{equation}\label{beta function to two loop order}
\alpha(\mu^2)=\frac{4\pi}{\beta_0}\left[\frac{1}{\log(\mu^2/\Lambda^2)} -\frac{\beta_1}{\beta_0^2}\frac{\log\log(\mu^2/\Lambda^2)}{\left(\log(\mu^2/\Lambda^2)\right)^2}\right]\,,
\end{equation}
where $\mu$ is the renormalization scale, $\beta_0=(22-4n_f)/3$,   $\beta_1=(136-64n_f)/3$, and $\alpha=g^2/(4\pi)$. We only consider a small number of Weyl adjoint fermions, $n_f < 5.5$, to preserve the asymptotic freedom.

An important variable is the Polyakov loop, or holonomy, defined as the path ordered exponent in the $\S^1$ direction:
\begin{equation}
\label{omega1}
\Omega(x)=Pe^{i\int_{\S^1} dx^3 A^3(x,x^3)}\,,
\end{equation}
where $x\in \R^3$. This quantity transforms under $x$-dependent gauge transformation as $\Omega\rightarrow U^{-1}\Omega U$, and hence the eigenvalues of $\Omega$ are gauge invariant. The eigenvalues of $\Omega$ are parameterized as:
\begin{equation}
\label{omega}
\Omega=\mbox{diag}\left(e^{2\pi i\mu_1},e^{2\pi i\mu_2}\right),~~\mu_1+\mu_2=0. 
\end{equation} 

The holonomy is called {\em trivial} if the vacuum expectation value $\langle\Omega\rangle$ belongs to one of the elements of the group center $\Z_2$. Center symmetry transformations can be thought of as  ``gauge" transformations periodic up to an element of the center $U(x, x^3 + L) = z U(x, x^3)$, with $z = \pm 1$. Under such transformations, the Polyakov loop transforms as:  ${\rm{tr}} \Omega \rightarrow z \; {\rm{tr} } \Omega$. Thus when the vacuum expectation value of $\langle \Omega \rangle$ is 
 an element of the center,  the center symmetry is broken. 
 
 On the other hand, the holonomy is often said to be  {\em confining} if $\mbox{Tr}\,\langle\Omega\rangle=0$ (this terminology stems from the fact that in thermal Yang-Mills theory  $\tr \langle \Omega\rangle \sim e^{- F_q/T}$, where $F_q$ is the free energy of a static quark, serves as the order parameter for the deconfinement transition). For us, a more appropriate terminology is to call such holonomies {\em center-symmetric}, as one can have such vacua, e.g., in the $N=4$ supersymmetric Yang-Mills theory---not a confining, but a conformal theory---on $\R^3 \times \S^1$  with periodic boundary conditions \cite{Unsal:2010qh}. 
 
In this paper, we will be interested exclusively in vacua with {\em center-symmetric} holonomy.
  Then, as we shall show below, at distance scales  much larger than the $\S^1$ size $L$ the coupling constant $g$ ceases to run. Further, taking $1/L\gg\Lambda$, we remain in the perturbative regime and  can reliably perform a Coleman-Weinberg analysis \cite{Coleman:1973jx} to study the breaking of center symmetry. Integrating out the heavy Kaluza-Klein modes along $\S^1$, with periodic boundary conditions for the 
 fermions---remembering that our $\S^1$ is a spatial, not a thermal circle---results in an effective potential for $\Omega$:
\begin{eqnarray}
\label{abc}
V_{\mbox{\scriptsize eff}}(\Omega)=(-1+n_f)\frac{2}{\pi^2L^4}\sum_{n=1}^{\infty}\frac{1}{n^4}|\mbox{tr}\, \Omega^n|^2\,.
\end{eqnarray} 
Since $\Omega(x)\equiv e^{iL A_3(x)}$, see eqn.~(\ref{omega1}), the action for the $x^3$-independent modes of the gauge field is:\footnote{To avoid confusion, we note that in subsequent sections we study the Euclidean theory and, for convenience, relabel the compact direction $x^4$ and the ``Higgs field" $A_4$.}
\begin{eqnarray}
\label{action1}
S_0=\int_{\R^{1,2}}\frac{L}{g^2}\mbox{tr}\left[ -\frac{1}{2}F_{\mu\nu}F^{\mu\nu}+\left(D_\mu A_3\right)^2  - {g^2\over 2}V_{eff}(A_3)+ 2 i g^2\bar \lambda ^I \left(\bar\sigma^\mu D_\mu - i \bar \sigma_3\left[A_3 \,\,,\right] \right)\lambda_I\right]\,.
\end{eqnarray} 
The minimum of the potential $V_{\mbox{\scriptsize eff}}$ (\ref{abc}) for $n_f > 1$ is located at (see also the equivalent form in eqn.~(\ref{higgspotential2})  below):
\begin{equation}\label{holonomy of fundamental field}
\langle \Omega \rangle=\left(
\begin{array}{cc}
e^{i\pi/2}&0\\
0&e^{-i\pi/2}
\end{array}\right)\,, \quad \mbox{or} \quad
\langle A_3\rangle \equiv \langle A_3^3 \rangle \; t^3   = {\pi \over L } \; t^3 \,,
\end{equation}
and hence the center symmetry is preserved since $\mbox{Tr}\, \langle \Omega \rangle=0$  \cite{Unsal:2006pj} . This is in contrast with the $n_f = 0$ 
case, which is equivalent to a finite temperature pure Yang-Mills theory, where the theory deconfines at sufficiently high temperature \cite{Gross:1980br}.
In the supersymmetric $n_f = 1$ case the Coleman-Weinberg potential vanishes and the center-symmetric vacuum is stabilized due to nonperturbative corrections \cite{Seiberg:1996nz,Aharony:1997bx,Davies:1999uw}.

In the vacuum (\ref{holonomy of fundamental field}), the $SU(2)$ gauge symmetry is broken by the Higgs mechanism down to $U(1)$ at the scale of $\langle A_3^3 \rangle = {\pi\over L}$. Because the Higgs is in the adjoint, two of its components are eaten by the gauge fields and become massive, with $m_W = {\pi\over  L}$. The third component of the Higgs field obtains mass due to the one-loop potential $V(A_3)$. A ``friendlier" version of  the potential  (\ref{abc}) as a function of the ``Higgs field" $A_3^3$ is given by:
\begin{equation}
\label{higgspotential2}
V_{\mbox{\scriptsize eff}}(A_3^3)= - {n_f - 1 \over 12 \pi^2} [A_3^3]^2 \left( {2 \pi \over L } - [A_3^3] \right)^2 + {\rm field \; independent},
\end{equation}
where $[A_3^3] = A_3^3$ mod(${2 \pi \over L}$), and is clearly (recall that $n_f > 1$) minimized at $\langle A_3^3\rangle ={\pi\over L}$. The ``Higgs" mass is then $m_H  = {g\over L} {\sqrt{  n_f -1   \over  3}}$ and the ratio ${m_H \over m_W} 
=   g   \sqrt{n_f -1  \over  3} $. This ratio will be of some interest to us in what follows, and we record it below:
\begin{equation}
\label{mhovermw}
{m_H\over m_W} = c g~,
\end{equation}
 with $c$ an order one number determined above.

In addition, two of the fermion components (the color space components of $\lambda$ that do not commute with $\langle A_3 \rangle$) acquire a mass $\sim 1/L$. Therefore, for distances $\gg L$, the $3D$ low energy theory is:
\begin{equation}
S=L\int_{\R^{1,2}}-\frac{1}{4g^2}F_{3,\mu\nu}^2+i\bar\lambda_{3}^I\bar\sigma^\mu\partial_{\mu}\lambda_{3,I}\,,
\end{equation}  
which describes  a non-interacting gauge field   and neutral fermions. Hence, the coupling $g$ ceases to run at $1/L \gg \Lambda$, as was anticipated above.  

Finally, since the gauge field $F_{3,\mu\nu}$ lives in $3D$, it carries only one degree of freedom. A duality transformation can be performed by first introducing  a Lagrange multiplier field $\sigma$ as 
$\delta S= {1\over 8 \pi} \int_{\R^3}  \sigma \epsilon_{\mu\nu\lambda} \partial^\mu F_{3}^ {\nu\lambda}$, to enforce the Bianchi identity, then  integrating over $F_{\mu\nu}$, and substituting $F_{3,\mu\nu} = - {g^2 \over 4 \pi L} \epsilon_{\mu\nu\lambda} \partial^\lambda \sigma$ into $S + \delta S$. The result is the Lagrangian in terms of the ``dual photon" field $\sigma$: 
\begin{equation}
\label{perturbativeSeff}
S= \int_{\R^{1,2}}{1\over 2 L} \left( g \over 4 \pi  \right)^2 \left(\partial \sigma\right)^2+ L \; i\bar\lambda_{3}^I\bar\sigma^\mu\partial_{\mu}\lambda_{3,I}\,~,
\end{equation}  
which captures the long-distance physics to all orders in perturbation theory and to leading order in the derivative expansion. We note for later use that with our normalization of magnetic charge ($Q_m = \pm 1$, etc., $ \int_{\R^3}    \epsilon_{\mu\nu\lambda} \partial^\mu F_{3}^{\nu\lambda} = 8 \pi Q_m$), the form of $\delta S$ implies that $\sigma$ is a compact scalar with period $2 \pi$.

\section{Instantons and instanton-monopoles}
\label{instantons}

\subsection{Instanton solutions}
\label{instantons2}

In the previous section, we sketched the perturbative treatment of the theory in a center-symmetric vacuum at energies $\ll 1/L$. However,    there is a more interesting story to tell thanks to nonperturbative effects. This follows from the existence of instanton  and monopole-instanton  solutions which describe the tunneling processes between different classical vacua (for reviews see \cite{Schafer:1996wv,Diakonov:2002fq,Vandoren:2008xg}). These are nontrivial topological solutions of finite Euclidean action $S_E$. Since the space is Euclidean, we can shuffle the indices such that $4$ refers  to the compactified $\S^1$ (which will occasionally also be called the ``Euclidean time"---even though we should keep in mind that for us $\S^1$ is a compactified spatial direction and the fermions are periodic), and $i,j$ are reserved for the infinite dimensions. 

 The complete information about the perturbative and nonperturbative dynamics of (\ref{basic lagrangian}) is encoded in the Euclidean version of the partition function:
\begin{equation}\label{Euclidean partition function}
Z=\int D[ A_{M}]D[\lambda_{I}]D[\bar\lambda_{I}]\exp[-S_E]\,,
\end{equation} 
where:
\begin{eqnarray}\label{Euclidean basic lagrangian}
S_E=\int_{\R^3\times \S^1}\frac{1}{4g^2}F_{MN}^aF_{MN}^a+i\bar \lambda_{a}^I \bar\sigma^M D_M^{ab}\lambda_{b\,I}\,,
\end{eqnarray}
where we use the notation of  \cite{Paik:2009iz} for the Euclidean space quantities (throughout the rest of the paper all quantities are understood to be in Euclidean space).

To find  instanton solutions, we first neglect the effect of the fermions. Then, the tunneling path is a solution with minimum Euclidean action connecting the vacua with different topological numbers. To find these solutions, it is more convenient to write the action in the form:
\begin{eqnarray}\label{pure Euclidean action}
S_E=\int_{\R^3\times \S^1}\frac{1}{4g^2}F_{MN}^aF_{MN}^a=\frac{1}{4g^2}\int_{\R^3\times \S^1}\left[\pm F_{MN}^a\tilde F_{MN}^a+\frac{1}{2}\left(F_{MN}^a\mp\tilde F_{MN}^a \right)^2\right]\,,
\end{eqnarray}
where $\tilde F_{MN}=\epsilon_{MNPQ}F_{PQ}/2$ is the dual field strength tensor. The action is at a minimum if $F_{MN}$ satisfies the (anti) self-dual condition:
\begin{equation}
F_{MN}^a=\pm\tilde F_{MN}^a\,.
\end{equation} 
It is straightforward  to check that these solutions satisfy the Euclidean equations of motion $D_{M}F_{MN}=0$. The topological charge (winding number or Pontryagin number) is given by:
\begin{equation}
Q_T=  \frac{1}{32\pi^2}\int_{\R^3\times \S^1} F_{MN}^a\tilde F_{MN}^a\,.
\end{equation}
Hence, for (anti) self-dual solution the action is $S_E=\frac{8\pi^2|Q_T|}{g^2}$, and the tunneling amplitude is $\sim e^{-S_E}=\exp(-\frac{8\pi^2|Q_T|}{g^2})$.   

According to Gross, Pisarski, and Yaffe \cite{Gross:1980br}, the classical configurations with a finite action can be classified by ({\it i.})  topological charge $Q_T$, {\it (ii.)} holonomy at infinity:
\begin{equation}
\langle \Omega\rangle =Pe^{i\int_{\S^1} dx^4A^4(x,x^4)}\bigg\vert_{x \rightarrow \infty}\,,
\end{equation}
and ({\it iii.})  magnetic charge, $Q_m$, or the flux of the chromo-magnetic field (shown below in the vacuum (\ref{holonomy of fundamental field})) at spatial infinity:
\begin{equation}
\int\limits_{\S^2_\infty} d^2 \vec{\sigma} \vec{B} = 4 \pi Q_m t^3\; , ~~~ \vec B=Q_m \frac{\vec r}{ |\vec r|^3}  t^3  \,.
\end{equation}

The solutions with  $|Q_T|=1$ and trivial holonomy are the well-known $SU(2)$  BPST (anti) self-dual instanton
 solution in non-compactified $4D$ Euclidean  space and are the leading tunneling   field configuration for zero temperature. The BPST solution was generalized by Harrington and Shepard \cite{Harrington:1978ve} to non-zero temperatures, $T$,  by compactifying one of the Euclidean dimensions on a circle of radius $L=1/T$. This ``caloron" solution  was obtained by performing a summation over infinite number of BPST images located periodically on the ``uncompactified" circle. The Harrington-Shepard configuration has trivial holonomy, $\mbox{Tr}\, \langle \Omega \rangle \ne 0$, and zero magnetic charge since the field strength falls off fast enough at spatial infinity; consequently, as we will see,  calorons with trivial holonomy do not lead to confinement. 
 
 In order to obtain a caloron solution with non-trivial holonomy, one has to ensure that $\langle A_4\rangle =\pi T\tau_3/2$, or a center-symmetric holonomy, $\mbox{Tr}\,\langle \Omega\rangle =0$. The field configuration that obeys this requirement was found by Kraan and van Baal \cite{Kraan:1998pm} using the Atiyah-Drinfeld-Hitchin-Manin (ADHM) construction \cite{Atiyah:1978ri}, and independently by Lee and Lu  \cite{Lee:1998bb} in the context of D-branes. The caloron with center-symmetric holonomy solution is lengthy, and it is not very illuminating to write it explicitly. However,  for a ``fat" enough caloron one can show that the solution can be broken down into a simple linear superposition of two constituent {\em monopoles} (more appropriately, these should be referred to as ``monopole-instantons", but we will occasionally call them monopoles).   As we shall show in subsequent sections, these monopole configurations are the  topological defects essential in understanding confinement in  $SU(N)$ QCD(adj) theories at small $\S^1$.

\subsection{Monopole-instanton solutions}
\label{monopoleinstantons}

Monopoles are static, ``time" ($x_4$) independent, and (anti) self-dual solutions of the Euclidean equations of motion $D_MF_{MN}=0$.\footnote{Here we follow the convention of \cite{Kraan:1998pm}. Our brief review of monopoles is heavily based on \cite{Diakonov:2009jq}. We stress again that $x^4$ is a compactified {\em spatial} direction. Hence, our referring to $F_{i4}$ as  ``electric" and $B_i$ as ``magnetic" below is only for convenience of speech (we will continue to use these terms, but keeping in mind the present remark, will usually surround them by quotation marks). Let us also note that the term ``magnetic" does, however,  have a physical interpretation in the case of a spatial $\S^1$ compactification: the monopole-instantons described here correspond to tunneling events between vacua of the 3d theory of different numbers of magnetic (i.e., $F_{12}$) flux quanta (when defined on a  finite spatial $T^2$ in, say, the $x_{1},x_2$ directions; these different flux vacua become degenerate in the limit of infinite $T^2$ size). }
 These solutions have finite action,  topological charges are possibly fractional, and possess non-trivial holonomy. Their ``chromo-electric" and ``chromo-magnetic" field strengths are
\begin{equation}
E_i^a=F_{i4}^a=D_{i}^{ba}A_4^b\,\quad B_i^a=\frac{1}{2}\epsilon_{ijk}F_{jk}^a\,,
\end{equation}
where $D^{ab}_i=\partial^i\delta^{ab}+\epsilon^{abc}A^c_i$, and $F_{MN}^a=\partial_{M}A_{N}^a-\partial_NA_M^a+\epsilon^{abc}A_{M}^bA_N^c$\,. In terms of these fields, the action (\ref{pure Euclidean action}) reads:
\begin{eqnarray}
\nonumber
S_E&=&\frac{L}{2g^2}\int d^3x\left[\left(E_i^a\right)^2+\left(B_i^a\right)^2 \right]=\frac{L}{2g^2}\int d^3x \left(E_i^a\mp B_i^a\right)^2\pm\frac{L}{g^2}\int d^3x E_i^aB_i^a\\
\label{action in terms of E and B}
&&\geq \pm \frac{L}{g^2} \int d^3 xE_i^a B_i^a \,.
\end{eqnarray}
The equality is satisfied for (anti) self-dual solutions:
\begin{equation}\label{self dual solution for monopole} 
F_{MN}^a=\pm {1 \over 2} \epsilon_{MNPQ}F_{PQ}^a \quad \mbox{or} \quad  E_i^a= \pm B_i^a \,.
\end{equation}
To find an explicit form of the monopole solution one uses the hedgehog ansatz 
\begin{equation}
A_{4}^a=n_a\frac{{\cal E}(r)}{r}\,,\quad A_{i}^a=\epsilon_{aij}n_j\frac{1-{\cal Z}(r)}{r}\,,
\end{equation}
in (\ref{self dual solution for monopole}), where $n_a=r_a/|\vec r|$ is a unit vector in the direction of $r_a$. Then, equating the corresponding terms, one obtains two equations in ${\cal Z}$ and ${\cal E}$. The numerical solution to the resulting equations was first found by Prasad and Sommerfield \cite{Prasad:1975kr}. Later, a simple analytic solution was found by Bogomolny \cite{Bogomolny:1975de}, see the following Section \ref{bpsmonopoles} for explicit formulae. 
This solution satisfies the (anti) self-dual condition $E_i^a=\mp B_i^a$ and its action is: 
\begin{equation}\label{second equation for monopole solution}
S_E=\frac{4\pi L}{g^2}\left|\int dr r^2 B_i^a B_i^a\right|=\frac{4\pi L v}{g^2}\,.
\end{equation}
These monopoles carry both ``electric" and ``magnetic" charges (and hence sometimes are also called dyons \cite{Diakonov:2009jq}, but we find that terminology confusing for our purposes). In the following we classify the four different monopole solutions depending on all possible signs of ``electric" and ``magnetic" charges.

\subsubsection{Bogomolny-Prasad-Sommerfield (BPS) monopoles}
\label{bpsmonopoles}

The BPS monopole solution is, in the regular (hedgehog) gauge:
\begin{eqnarray}
\nonumber
A_4^a&=&\mp n_av{\cal P}(vr)\,,\\
\label{all hedgehog gauge}
A_{i}^a&=&\epsilon_{aij}n_j\frac{1-{\cal A}(vr)}{r}
\end{eqnarray}
where the upper sign corresponds to the self-dual (BPS), and the lower one to the anti self-dual ($\overline{\mbox{BPS}}$) solution, where: 
\begin{eqnarray}
\nonumber
{\cal P}(x)&=&\coth x-\frac{1}{x}\stackrel{x\rightarrow \infty}{\rightarrow}1-\frac{1}{x}\,,\\
{\cal A}(x)&=&\frac{x}{\sinh x}\stackrel{x\rightarrow \infty}{\rightarrow}{\cal O}\left(xe^{-x} \right)\,.
\end{eqnarray}
The magnetic field strength is:
\begin{eqnarray}
B_i^a=\left(\delta_{ai}-n_an_i\right){\cal F}_1(vr)+n_an_i{\cal F}_2(vr)\,,
\end{eqnarray}
with:
\begin{eqnarray}
\nonumber
{\cal F}_1(vr)&=&\frac{v^2}{\sinh vr}\left(\frac{1}{vr}-\coth vr \right)\stackrel{r\rightarrow \infty}{\rightarrow}v^2{\cal O}\left(e^{-vr}\right)\,,\\
\label{F1 and F2}
{\cal F}_2(vr)&=&\frac{v^2}{\sinh ^2 vr}-\frac{1}{r^2}\stackrel{r\rightarrow \infty}{\rightarrow}-\frac{1}{r^2}\,.
\end{eqnarray}
The electric field is $E_i^a=B_i^a$ for the BPS and $E_i^a=-B_i^a$ for the $\overline{\mbox{BPS}}$. 

If there is more than one monopole in the vacuum, then it becomes impossible to add them up since $A_4^a$ have different values at asymptotic infinity. To overcome this problem, i.e, to achieve the same value of $A_4^a$ at infinity, we gauge-transform the solutions 
$A_\mu\rightarrow UA_\mu U^{\dagger}+iU\partial_\mu  U^{\dagger}$, where $U$ is any of the matrices:
\begin{eqnarray}
\nonumber
S_+(\theta,\phi)&=&e^{-i\phi\tau^3/2}e^{i\theta\tau^2/2}e^{i\phi\tau^3/2}\,,\\
S_-(\theta,\phi)&=&e^{i\phi\tau^3/2}e^{i(\pi+\theta)\tau^2/2}e^{i\phi\tau^3/2}\,,
\end{eqnarray}
where $\theta$ and $\phi$ are the spherical angles.\footnote{Our convention for spherical coordinates is $\vec r= (x^1,x^2,x^3)= \left( r\sin\theta\cos\phi\,, r\sin\theta\sin\phi\,, r\cos\theta\right)$. }
This gauge transformation brings the solution to the  ``stringy gauge." We use $S_-$ and $S_+$, respectively, to gauge-transform the BPS and $\overline{\mbox{BPS}}$ solutions. Using the identities $S_+(\vec n \cdot \vec\tau)S^{\dagger}_+=\tau^3$ and $S_-(\vec n \cdot \vec\tau)S^{\dagger}_-=-\tau^3$ we obtain:
\begin{equation}
A_4^{\mbox{\scriptsize BPS}\,,\overline{\mbox{\scriptsize BPS}}}\equiv A_4^{a\,\mbox{\scriptsize BPS}\,,\overline{\mbox{\scriptsize BPS}}}\frac{\tau^a}{2}=v{\cal P}(vr)\frac{\tau^3}{2}\stackrel{r\rightarrow \infty}{\rightarrow}v\frac{\tau^3}{2}\,,
\end{equation}
and: 
\begin{eqnarray}
A_i^{\mbox{\scriptsize BPS}\,,\overline{\mbox{\scriptsize BPS}}}=\left\{ \begin{array}{l} A_r=0\\
A_\theta=\frac{{\cal A}(vr)}{2r}\left(\pm\tau^1\sin\phi+\tau^2\cos\phi\right)\\
A_\phi=\frac{{\cal A}(vr)}{2r}\left(\pm\tau^1\cos\phi-\tau^2\sin\phi\right)\pm\tau^3\frac{1}{2r}\tan\frac{\theta}{2}
   \end{array} \right.\,,
\end{eqnarray}
with holonomy:
\begin{equation}
\Omega^{\mbox{\scriptsize BPS}\,,\overline{\mbox{\scriptsize BPS}}}=\left(\begin{array}{cc} e^{iLv/2}&0\\0&e^{-iLv/2}  \end{array} \right)\,.
\end{equation}
The magnetic field in the stringy gauge is given by: 
\begin{eqnarray}
B_i^{\mbox{\scriptsize BPS}\,,\overline{\mbox{\scriptsize BPS}}}=\left\{ \begin{array}{l} B_r=\mp\frac{{\cal F}_2(r)}{2}\tau^3\stackrel{r\rightarrow \infty}{\rightarrow}\pm\frac{1}{r^2}\frac{\tau^3}{2}\\
B_\theta=\frac{{\cal F}_1(r)}{2}\left(\mp\tau^1\cos\phi+\tau^2\sin\phi\right)\stackrel{r\rightarrow \infty}{\rightarrow}v^2{\cal O}(e^{-vr})\\
\label{magnetic field for the BPS monopole}
B_\phi=\frac{{\cal F}_1(r)}{2}\left(\pm\tau^1\sin\phi+\tau^2\cos\phi\right)\stackrel{r\rightarrow \infty}{\rightarrow}v^2{\cal O}(e^{-vr})
   \end{array} \right.\,,
\end{eqnarray}
where the upper (lower) sign is for BPS ($\overline{\mbox{BPS}}$). The electric field is $E_i^{\mbox{\scriptsize BPS}}=B_i^{\mbox{\scriptsize BPS}}$, and $E_i^{\overline{\mbox{\scriptsize BPS}}}=-B_i^{\overline{\mbox{\scriptsize BPS}}}$. We see that the $\theta$ and $\phi$ components have non-Abelian fields but die away quickly outside the monopole cores, i.e., for $r\gg 1/v$. Hence, $1/v$ is interpreted as the monopole core size. On the other hand, the radial component behaves like an Abelian field as it has only the third color component.

\subsubsection{``Kaluza-Klein" (KK) monopoles}
\label{kkmonopoles}

Another class of solutions is known in the literature as ``Kaluza-Klein" (KK) or ``twisted" monopoles. This includes both the self-dual KK and the anti self-dual $\overline {\mbox{KK}}$ solutions. 

The existence of the twisted monopole-instantons is only possible because  the ``Higgs field" $\sim e^{i L A_4}$ is  compact.
We note that the existence of extra monopole solutions in theories with compact Higgs fields has been noted, but not pursued,  earlier, in the context of maximal abelian projection  \cite{Kronfeld:1987vd}. The advent of $D$-branes  greatly helped  the study of the twisted monopole-instantons, as they appear rather naturally in  string theory brane constructions  \cite{Lee:1997vp}. The high degree of supersymmetry typical of these constructions does not preclude inferring the existence of bosonic classical solutions also in theories without supersymmetry.

Compactness of the ``Higgs field" means that the expectation value of $A_4$  of eqn.~(\ref{holonomy of fundamental field}) is a periodic variable $\langle A_4^3\rangle  \equiv \langle A_4^3\rangle + {2 \pi \over L}$ and vacua with values of $\langle A_4^3\rangle$ differing by an integer number times $2 \pi \over L$ are equivalent. The equivalence can be seen by first noting that $A_4$ always enters the lagrangian in the combination $\partial_4 + i A_4$ and then either relabeling the
Kaluza-Klein (``Matsubara") frequencies or (equivalently)  performing a ``large" gauge transformation, $A_4^3 \rightarrow A_4^3 + \partial_4 \left[{2 \pi k \over L}  x_4\right]$, to absorb any part of the expectation proportional to an integer $\times 2 \pi/L$.

To obtain the twisted solutions one  starts with a ``static" $x_4$-independent BPS monopole solution of unit charge (i.e. the solutions considered in Section \ref{bpsmonopoles}) in a vacuum where the expectation value $v$ of $A_4$ is shifted by any $2 \pi k \over L$, $k \in \Z$. These ``static" solutions will have action proportional to $|{2 \pi k \over L} + v|$, as per (\ref{second equation for monopole solution}). One then ``gauge" transforms the expectation value of $A_4$ back to $v$, which introduces an $x_4$-dependent ``twist" of the solution.
In this manner one obtains  an infinite tower of ``twisted" instanton-monopole solutions of action ${4 \pi L \over g^2} |  {2 \pi k \over L} + v|$ (for magnetic charge $\pm 1$). 

In this paper, we will be only concerned with the lowest action ``twisted" solutions, which are the only ones relevant at sufficiently small $L$. 
These are obtained from the equations in the previous section by replacing $v\rightarrow \frac{2\pi}{L}-v$. One first transforms from the hedgehog to the stringy gauge with the help of $S_+$ (for KK, and we use the upper sign in eq.~(\ref{all hedgehog gauge})) and $S_-$ (for $\overline {\mbox{KK}}$, and we use the lower sign in eq. (\ref{all hedgehog gauge})). As a result one finds $A_4(\infty)=(-2\pi/L+v)\tau^3/2$. To put the asymptotic form of $A_4$ in the same form as BPS and $\overline{\mbox{BPS}}$ we use $S^1$-dependent matrix (a  ``gauge" transformation periodic up to the center of the gauge group): 
\begin{equation}
U=e^{i\; {\pi x^4\over L}\tau^3}\,.
\end{equation}
This results in the following fields for the KK and  $\overline {\mbox{KK}}$ monopoles in the stringy gauge:
\begin{eqnarray}
A_4^{\mbox{\scriptsize KK}\,,\overline {\mbox{\scriptsize KK}}}=\left[(-2\pi/L+v){\cal P}\left(|2\pi/L-v|r\right)+2\pi/L \right]\frac{\tau^3}{2}\stackrel{r\rightarrow \infty}{\rightarrow}\frac{v\tau^3}{2}\,,
\end{eqnarray}
 hence, $\Omega^{\mbox{\scriptsize KK}\,,\overline{\mbox{\scriptsize KK}}}=\Omega^{\mbox{\scriptsize BPS}\,,\overline{\mbox{\scriptsize BPS}}}$. The vector potential and magnetic field read:
\begin{eqnarray}
A_i^{\mbox{\scriptsize KK}\,,\overline{\mbox{\scriptsize KK}}}=\left\{ \begin{array}{l} A_r=0\\
A_\theta=\frac{{\cal A}(|2\pi/L-v|r)}{2r}\left[\tau^1\sin\left(\frac{2\pi x_4}{L}\mp\phi\right)+\tau^2\cos\left(\frac{2\pi x_4}{L}\mp\phi\right)\right]\\
A_\phi=\frac{{\cal A}(|2\pi/L-v|r)}{2r}\left[\mp\tau^1\cos\left(\frac{2\pi x_4}{L}\mp\phi\right)\pm\tau^2\sin\left(\frac{2\pi x_4}{L}\mp\phi\right)\right]\mp \tau^3\frac{1}{2r}\tan\frac{\theta}{2}
   \end{array} \right.\,,
\end{eqnarray} 
\begin{eqnarray}
B_i^{\mbox{\scriptsize KK},\overline{\mbox{\scriptsize KK}}}=\left\{ \begin{array}{l} B_r=\pm\frac{{\cal F}_2(|2\pi/L-v|r)}{2}\tau^3\stackrel{r\rightarrow \infty}{\rightarrow}\mp\frac{1}{r^2}\frac{\tau^3}{2}\\
B_\theta=\frac{{\cal F}_1(|2\pi/L-v|r)}{2}\left[\pm\tau_1\cos\left(\frac{2\pi x_4}{L}\mp\phi\right)\mp\tau_2\sin\left(\frac{2\pi x_4}{L}\mp\phi\right) \right]\\
B_\phi=\frac{{\cal F}_1(|2\pi/L-v|r)}{2}\left[\tau_1\sin\left(\frac{2\pi x_4}{L}\mp\phi\right)+\tau_2\cos\left(\frac{2\pi x_4}{L}\mp\phi\right) \right]
   \end{array} \right.\,.
   \label{explicit time dependence in kk monopole}
\end{eqnarray}
The electric field is $E_i^{\mbox{\scriptsize KK}}= B_i^{\mbox{\scriptsize KK}}$, and $E_i^{\overline{\mbox{\scriptsize KK}}}= -B_i^{\overline{\mbox{\scriptsize KK}}}$. Notice that the $\theta$ and $\phi$ field components are dependent on the $S^1$ coordinate. Hence, the KK monopoles do not exist in a true $3D$ theory.

\begin{table}
\centerline{
\begin{tabular}{|c  |c  |c  |c| c|}
	\hline
	 & BPS& $\overline{\mbox{BPS}}$ & $\mbox{KK}$ & $\overline{\mbox{KK}}$ \\\hline
	 \mbox{``electric" charge} $Q_e$ & $+$ & $+$ & $-$ & $-$\\\hline
	\mbox{``magnetic" charge} $Q_m$ & $+$ & $-$ & $-$ & $+$ \\\hline
	$S_E$  &$\frac{4\pi L v}{g^2}$ & $\frac{4\pi L v}{g^2}$ & $\frac{8\pi^2}{g^2}-\frac{4\pi L v}{g^2}$ & $\frac{8\pi^2}{g^2}-\frac{4\pi L v}{g^2}$\\\hline
	$Q_T$ & $   \frac{vL}{2\pi}$ & $ -\frac{vL}{2\pi}$ & $  1-\frac{vL}{2\pi}$ & $  \frac{vL}{2\pi}-1$\\\hline
\end{tabular}
}
\caption{ ``Electric" and ``magnetic" charges, action, and topological charge for the fundamental  monopoles, i.e., the ones of least action  in the center-symmetric vacuum $v L = \pi$.}
\label{summary monopoles}
\end{table}

To conclude the presentation of the various solutions, 
 in Table~\ref{summary monopoles} we give a  summary of the different fundamental monopole solutions (the ones of lowest action in the center-symmetric vacuum $v = \pi/L$) and their charges. These will be important later when discussing their interactions. When placed a distance $r$ apart, $r v \gg 1$, two ``monopoles" with ``electric" and ``magnetic" charges $Q_e^{1,2}$ and $Q_m^{1,2}$, respectively, have an interaction ``energy" (i.e., Euclidean action, see eq.~(\ref{final expression for Sint})): 
\begin{equation}
\label{interaction1}
S_{int.}^{1-2} = {4\pi L \over g^2} \; {Q_m^{1} Q_m^2 - Q_e^1 Q_e^2  \over  r}~.
\end{equation}
The negative sign in front of the ``electric" interaction is due to the fact that this is really a 3d massless scalar ($A_4^3$, in the absence of the loop-generated potential (\ref{higgspotential2}), is massless) exchange, which is attractive for equal scalar charges. We note that this interaction will not be important for us---the reason is that  $A_4$ is gapped in our theory due to the one-loop effective potential (\ref{higgspotential2}) and our monopole-instanton solutions are only approximately BPS.

\subsubsection{Relation to  BPST-instantons with nontrivial holonomy} 
\label{calorons}

As already mentioned, 
instantons with nontrivial holonomy were found by Kraan and van Baal \cite{Kraan:1998pm}, and independently by Lee and Lu \cite{Lee:1998bb}. At large separation, these instantons can be viewed as a combination of BPS and KK monopoles \cite{Kraan:1998sn}. Let $r_{12}$ be the monopole separation which is given by $r_{12}=|\vec x_1-\vec x_2|$, where $x_1$ and $x_2$ are the positions of the two monopoles. Also, let $v$ and $\bar v=2\pi/L-v$ be the inverse core sizes of the BPS and KK monopoles respectively. When the monopoles are far apart, i.e. $r_{12} \gg 1/v$, and $r_{12} \gg 1/\bar v$, the instanton solution behaves like two independent and ``static" objects, the constituent monopoles. However, as the monopoles merge, the configuration develops ``time" dependence. This happens because the KK monopole has  built in time-dependence as evident from (\ref{explicit time dependence in kk monopole}). The instanton action is $S_I=S_{\mbox{\scriptsize BPS}}+S_{\mbox{\scriptsize kk}}=8\pi^2/g^2$, its topological charge is $Q_T(I)=Q_T(\mbox{BPS})+Q_T(\mbox{KK})=1$, and it carries zero magnetic charge.

 This concludes our review of finite-action solutions with nontrivial holonomy on $\R^3 \times \S^1$.

\section{Dynamics of  confinement in QCD(adj)}
\label{dynamicsofconfinement}
We now go back to the dynamics of the QCD(adj) theory, described in Section 1, at small-$\S^1$ and consider the role of the various nonperturbative topological excitations discussed above.  We wish to find out how the all-loop perturbative long-distance effective lagrangian, 
eqn.~(\ref{perturbativeSeff}) is modified by nonperturbative effects. 
\subsection{The vacuum of the small-$\S^1$ regime of QCD(adj) theories } 
\label{vacuumatsmalll}

As the integer topological charge (anti)instantons described above are magnetically neutral, they do not contribute to  confinement in the small-$\S^1$ regime of QCD(adj) theories. This is because (see also the following Sections) confinement is associated with the generation of a mass gap of the dual photon field---to which only magnetically charged objects can contribute (recall that $\sigma$ is sourced by magnetic charge). 

 The other topological solutions, the fundamental BBS and KK monopoles, have fermionic zero modes (for the relevant index theorems, see \cite{Nye:2000eg, Poppitz:2008hr}) and hence also do not generate mass  for the dual photon.
However, as was shown in \cite{Unsal:2007jx}, BPS-$\overline{\mbox{KK}}$ molecules (bions)  will generate a mass term for the dual photon and lead to confinement of electric charges. Thus, the vacuum of these theories will be dominated by such nonperturbative ``bion molecule" configurations.

Comparing the holonomy of the different monopoles to the analysis of Section 1, see equation (\ref{holonomy of fundamental field}),  we find $v={\pi\over L}$. Hence, the Euclidean action $S_E$,  topological charge $Q_T$, and magnetic charge $Q_m$, of the various solutions are: 
\begin{eqnarray}
S_E(\mbox{BPS},\overline{\mbox{BPS}},\mbox{KK},\overline{\mbox{KK}})&=&{4\pi^2 \over g^2} ~,\nonumber \\
Q_T (\mbox{BPS},\mbox{KK})&=&{1\over 2}~ , \nonumber \\
 Q_T (\overline{\mbox{BPS}},\overline{\mbox{KK}})&=&-{1\over 2} ~, \\
 Q_m (\mbox{BPS},\overline{\mbox{KK}}) &=& 1 ~,\nonumber \\
  Q_m (\overline{\mbox{BPS}},\mbox{KK})&=& -1 \nonumber ~.
 \end{eqnarray}
  Finally, since in QCD(adj) $A_4$ is massive, with  mass $\sim g/L$, see eqn.~(\ref{mhovermw}),  the electric field $E_i^a=D_{i}^{ba}A_4^b$ is gapped and its  contribution to the monopole interactions can be neglected in the following analysis.\footnote{This is not so in the $n_f = 1$ supersymmetric case. The appearance of mass gap of $A_4$ in the supersymmetric case is due to nonperturbative effects, which can be loosely described as due to magnetically neutral  $\mbox{BPS}$-$\overline{\mbox{BPS}}$ and
$\mbox{KK}$-$\overline{\mbox{KK}}$ ``molecules."}
 
The magnetic bions (anti-bions) are bound states of BPS and $\overline{\mbox{KK}}$ (their ``antiparticle" $\overline{\mbox{BPS}}$ and KK) configurations. These composites are not solutions to the BPS equations, and do not correspond to (anti)self-dual solutions. However, they are permitted by symmetries and turn out to be stable quantum mechanically. Although bions do not carry a net topological charge, $Q_{T}(\mbox{bion})=Q_T( \overline{\mbox{KK}})+Q_T( \mbox{BPS})=0$, they carry a net magnetic charge $Q_{m}(\mbox{bion})=Q_m( \overline{\mbox{KK}})+Q_m( \mbox{BPS})=2$, and $Q_{m}(\overline{\mbox{bion}})=Q_m( \mbox{KK})+Q_m( \overline{\mbox{BPS}})=-2$. Therefore, the electromagnetic interaction between the constituents of these composites is always repulsive. However, due to the exchange of fermionic zero modes between the monopoles, an attractive force overcomes the Coulomb repulsion resulting in stable molecular configurations. 

Since  bions carry a net magnetic charge, they play a crucial role in confinement in QCD(adj)---the same role played by magnetic monopoles in the $3D$ Polyakov model \cite{Polyakov:1976fu}. It is the purpose of this work to examine this picture more closely by working out its fine details.

\subsection{Bion plasma and confined phase}
\label{bionplasma}

To this end we recall the Euclidean partition function (\ref{Euclidean partition function}) (omitting the subscript $E$ to reduce notational clutter):
\begin{eqnarray}
\nonumber
Z&=&\int D[ A_{M}]D[\lambda_{I}]D[\bar\lambda_{I}]\exp[-S_E]\\
&=&\sum_{\mbox{\scriptsize config}}\int \left[D A_{M} \right]\exp\left[-\int_{\R^3\times \S^1}\frac{1}{4g^2}F_{MN}^{a\,2}\right]\prod_f^{n_f}\mbox{Det}\left(i\bar\sigma^{M}D^{\mbox{\scriptsize cl}}_M\right)\,,
\end{eqnarray}
where $D_{M}^{\mbox{\scriptsize cl}}=\partial_M +\left[A_M^{\mbox{\scriptsize cl}},\, \right]$ is the covariant derivative computed in the classical background, and the sum includes both perturbative and nonperturbative contributions to the vacuum. 

We are interested only in the nonperturbative configurations that lead to confinement. Namely, we expand the theory about bion and anti-bion configurations. Now we look for the minimum of $S_E$ under the condition that there are several bions and anti-bions, as their contribution will be important in the limit of infinite-dimensional 3-space. Then the solution will be given by a linear superposition of the bion fields provided that these bions are far apart. They will interact due to the long range magnetic force. Remembering that each bion (anti-bion) carries  $+2$ ($-2$) units of charge, the partition function  of the {\it dilute gas} of bions and anti-bions will be, using eqn.~(\ref{interaction1}) for the magnetic interaction between two bions:
\begin{equation}
\label{zgas1}
Z_{\mbox{ \scriptsize bion gas}}=\sum_{N_{\pm}, q_a=\pm 2 }\frac{Z_{\mbox{\scriptsize bion}}^{N_+ + N_-}}{N_+! N_-!}\prod_j^{N_+ + N_-} \frac{d^3R_j}{L^3} \exp\left[-\frac{4\pi L}{g^2 }\sum_{a >  b}\frac{q_aq_b}{|\vec R_a-\vec R_b|}\right]\,.
\end{equation}
Here, $R_i$ are the locations of the (anti)bions and $Z_{\mbox{\scriptsize bion}}$ is a single (anti)bion molecular partition function:\footnote{$Z_{bion}$ appearing in this Section, in particular in eqn.~(\ref{photon mass}), has  the integration over the  center of mass of the bion factored out,  as it is included in the $d^3 R_j$ integrals. $Z_{bion}$ studied in the following Sections, on the other hand, includes this integration. We use the same letter to denote the two partition functions and hope that confusion will be avoided. }
\begin{equation}\label{partition function of the bion}
Z_{\mbox{\scriptsize bion}}=\int_{\mbox{\scriptsize BPS}-\overline{\mbox{\scriptsize KK}}} \left[D A_{M} \right]\exp\left[-\int_{R^3\times S^1}\frac{1}{4g^2}F_{MN}^{a\,2}\right]\prod_f^{n_f}\mbox{Det}\left(i\bar\sigma^{M}D_M^{\mbox{\scriptsize cl}}\right)\,.
\end{equation}
 In order to simplify  the partition function of eqn.~(\ref{zgas1}) one introduces an auxiliary field $\sigma$ to find \cite{Polyakov:1976fu}:
\begin{eqnarray}
\label{zdual}
\nonumber
Z_{\mbox{ \scriptsize bion gas}} &\sim&  \int D\sigma \exp\left[-{1 \over 2 L} \left( g \over 4 \pi\right)^2\int d^3 x \left(\nabla \sigma\right)^2 \right] \times \\
&& \; \; \times \sum_{N_{\pm}, \{q_a=\pm 2\}}\frac{Z_{\mbox{\scriptsize bion}}^{N_+ + N_-}}{N_+! N_-! }\int {d^3R_1...d^3R_{N_++N_-} \over L^{3(N_+ +N_-)}} \exp\left[ i\sum_{a=1}^{N_+ + N_-}q_{a}\sigma(\vec R_a)\right]\nonumber \\
&=&\int D \sigma \exp\left[-{1 \over 2 L} \left( g \over 4 \pi\right)^2\int d^3 x \left[\left(\nabla \sigma\right)^2 -{{\cal M}^2 \over 2}\cos 2 \sigma \right] \right]~,
\end{eqnarray}
where:
\begin{equation}\label{photon mass}
{\cal M}^2=  \frac{8 \left(4\pi\right)^2Z_{\mbox{\scriptsize bion}}(g)}{ g^2L^2}\,.
\end{equation}
The (``Debye-H\" uckel")  mean field $\sigma$ is sourced by magnetic charge 
and can be interpreted as the dual photon of eqn.~(\ref{perturbativeSeff}) (the 
determinant of the laplacian omitted in the transition between (\ref{zgas1}) and (\ref{zdual}) cancels the free dual photon partition function; clearly, if instantons are ignored, the partition function of the theory can be written as a path integral over the  fields appearing in the effective action (\ref{perturbativeSeff})).  
The mass $\cal{M}$ of eqn.~(\ref{photon mass}) will be henceforth referred to as the ``bion-induced mass of the dual photon." 

The partition function in the dual form (\ref{zdual})  incorporates the leading nonperturbative effects due to instantons and is  useful   to compute gauge invariant expectation values probing the long-distance dynamics of the theory. First and foremost, Wilson loops can be used to deduce the linear long distance potential between two heavy electric test charges. For a large rectangular surface $\Sigma\subset R^3$ with boundary $C=\partial \Sigma$ and area $r\times T$, the expectation value of the Wilson operator $W(C)$ is given, up to  ${\cal O}(1)$ numbers, by:
\begin{equation}
\lim\limits_{T\rightarrow \infty}  \frac{1}{T}\log\left\langle W(C)\right\rangle\sim \gamma r\,,
\end{equation}
where (see \cite{Polyakov:1976fu} for more detail; we note also the review \cite{Kogan:2002au}):
\begin{equation}
\label{tension}
\gamma \sim {g^2\over L} {\cal M}\,.
\end{equation}
This implies that there exists an electric string between the test charges with tension $\gamma$, essentially determined by the mass of the dual photon.
One can also calculate the correlation function of the magnetic field $B_i=\epsilon_{ijk}F_{jk}^3/2$ in the background 
of the bion plasma to find:
\begin{equation}
\label{bpole}
\left\langle B_i(k)B_j(-k)\right\rangle\sim\left\langle \partial_i \sigma (k)\partial_j \sigma(-k)\right\rangle \sim \frac{k_ik_j}{{\cal M}^2+k^2}\,.
\end{equation}
Again, the dual-photon mass ${\cal M}$ (or inverse Debye screening length), which characterizes the size of flux tubes in the confining phase,  appears in the correlator. 

The magnetic bion mechanism of confinement discussed here is a generalization of the Polyakov  3D Debye-screening mechanism to a locally 4D theory. Instead of monopoles, the objects responsible for the magnetic screenings are composites
of self-dual monopoles and twisted anti-self-dual monopoles. 
One of our main aims is to study in more detail the scales involved and to find the dependence of the dual photon mass on the circle size $L$. Hence, it is important to track the dependence of $Z_{\mbox{\scriptsize bion}}$ on the coupling constant $g$, as we do in the following sections.

\subsection{Bion structure}
\label{bionstructure}

The analysis of the previous section assumed the existence of stable molecular bions. In the following, it will be our main task to investigate  the microscopic structure of these bions and calculate their partition function  $Z_{\mbox{\scriptsize bion}}$.

In order to perform the integral over $A_M$ in (\ref{partition function of the bion}), we first decompose the fields into a background part and quantum fluctuations $A_M=A_{M}^{\mbox{\scriptsize cl}}+A_{M}^{\mbox{\scriptsize qu}}$, where $A_{M}^{\mbox{\scriptsize cl}}(\vec x,\vec x_{\mbox{\scriptsize BPS}},\vec x_{\overline{\mbox{\scriptsize KK}}})=A_M^{\mbox{\scriptsize BPS}}\left(\vec x-\vec x_{\mbox{\scriptsize BPS}}\right)+A_M^{\overline{\mbox{\scriptsize KK}}}\left(\vec x-\vec x_{\overline{\mbox{\scriptsize KK}}}\right)$ is the classical background field of the BPS-$\overline{\mbox{KK}}$ pair. Before expanding, we also have to fix the gauge and introduce ghosts $c$ and anti-ghosts $b$. We choose the background gauge condition $D_{M}^{\mbox{\scriptsize cl}}A_{M}^{\mbox{\scriptsize qu}}=0$. Hence, the gauge fixing term is ${\cal L}_{\mbox{\scriptsize fix}}=-\mbox{tr}\left(D_M^{\mbox{\scriptsize cl}} A_M^{\mbox{\scriptsize qu}}\right)^2/g^2$, and the ghost action is ${\cal L}_{\mbox{\scriptsize ghost}}=-b^a\left(D^{\mbox{\scriptsize cl}}_MD_M^{\mbox{\scriptsize cl}} \right)c^a$.

The action, expanded to quadratic order in quantum fields, takes the form:
\begin{eqnarray}
\label{interaction}
\nonumber
S_E\left(\mbox{bion}=\mbox{BPS}+\overline{\mbox{KK}}\right)&=&S_E\left(\mbox{BPS}\right)+S_E\left(\overline{\mbox{KK}}\right)\\
\nonumber
&&+\underbrace{\frac{L}{g^2}\int d^3 x \vec B^{\overline{\mbox{\scriptsize KK}}}\left(\vec x-\vec x_{\overline{\mbox{\scriptsize KK}}}\right)\cdot \vec B^{\mbox{\scriptsize BPS}}\left(\vec x-\vec x_{\mbox{\scriptsize BPS}}\right)}_{\equiv {\rm \large S_{int}}}\\
\nonumber
&&+\underbrace{\frac{L}{g^2}\int d^3 x \vec E^{\overline{\mbox{\scriptsize KK}}}\left(\vec x-\vec x_{\overline{\mbox{\scriptsize KK}}}\right)\cdot \vec E^{\mbox{\scriptsize BPS}}\left(\vec x-\vec x_{\mbox{\scriptsize BPS}}\right)}_{\sim {\large 0}\, \mbox{as the electric field is gapped in QCD(adj)} }\\
&&-\frac{1}{g^2}\mbox{tr}\int d^4x\left[ A_M^{\mbox{\scriptsize qu}}M_{MN}A_N^{\mbox{\scriptsize qu}}+2b M^{gh}c \right]\,,
\end{eqnarray}
where $M^{gh}=\left(D^{\mbox{\scriptsize cl}}\right)^2$ and $M_{MN}=\left(D^{\mbox{\scriptsize cl}}\right)^2\delta_{MN}+2F^{\mbox{\scriptsize cl}}_{MN}$, and we also used (\ref{action in terms of E and B}). 

For our purposes it will be important to take into account the effect of the ``Higgs" potential $V(A_4)$ in eqn.~(\ref{action1}) on the solution. The fact that the potential is non-vanishing makes the solution not exactly BPS---but the fact that we work for small $g$ means that the ratio of Higgs to $W$-boson mass is small: $m_H/m_W \sim g$, see the discussion around eqn.~(\ref{higgspotential2}). This modifies the action of a single monopole at small $\beta = m_H/m_W$ as follows:
\begin{equation}
\label{higgscorrection}
S_{monopole}(m_H) = S_{BPS}\left( 1 + {1 \over 2} \beta + {1 \over 2} \beta^2 \ln \beta + c_3 \beta^2 + ...\right) ~,
\end{equation}
where $S_{BPS}$ denotes the action of a monopole in the $m_H = 0$ BPS limit.
The leading small-$\beta$ term has been calculated a long time ago  \cite{Kirkman:1981ck} and will be sufficient for our purposes, since we work at weak coupling and $\beta = c g$, see eqn.~(\ref{mhovermw}) (for completeness, we note the recent  calculation\footnote{The asymptotic expansion (\ref{higgscorrection}) is only valid \cite{Forgacs:2005vx} for $\beta \ll 10^{-3}$; see the further remarks in Section \ref{dualphotonmass}.} of $c_3$   \cite{Forgacs:2005vx}). Note that this result  applies equally well to KK monopoles, since they are obtained from static BPS solutions,  for which (\ref{higgscorrection}) is correct,  and the monopole action is gauge invariant.

Thus, since $S_E\left(\mbox{BPS}\right)+S_E\left(\overline{\mbox{KK}}\right) \approx {8\pi^2 \over g^2}( 1 + c g)$, the partition function (\ref{partition function of the bion}) takes the form:
\begin{equation}
Z_{\mbox{\scriptsize bion}}= \exp[{8\pi^2 \over g^2}( 1 + c g)]\exp[-S_{int}]\mbox{Det}\left(M_{MN}\right)^{-1/2}\mbox{Det}\left(M^{gh}\right)\prod_f^{n_f}\mbox{Det}\left(i\bar\sigma^{M}D_M\right)\,.
\end{equation}
However, the operator $M_{MN}$ has zero modes that need to be singled out and treated separately to avoid unphysical infinities. These zero mode solutions take the form:
\begin{equation}
Z_M^{(I)}=\frac{\partial A_M^{\mbox{\scriptsize cl}}}{\partial \gamma_I}+D^{\mbox{\scriptsize cl}}_M \Lambda^I\,,
\end{equation} 
where $I$ runs over the set of collective coordinates $\gamma_I$, and the gauge parameters $\Lambda^I$ are chosen such that $Z_M^{(I)}$ obeys  the background gauge condition, i.e. $D_M^{\mbox{\scriptsize cl}} Z_M^{(I)}=0$. In our case, there is a total of $8$ collective coordinates, $4$ for each monopole. These collective coordinates are divided into translation $ x_{\mbox{\scriptsize BPS}}^i,x_{\overline{\mbox{\scriptsize KK}}}^i$, where $i=1,2,3$,  and global $U(1)$ rotations $U(\theta_{\mbox{\scriptsize BPS}}),U(\theta_{\overline{\mbox{\scriptsize KK}}})$  for the BPS and $\overline{\mbox{KK}}$ monopoles respectively.  Defining the norms-squared  matrix ${\cal U}^{IJ}$, which can be interpreted as the metric on the moduli space of collective coordinates:
\begin{equation}
{\cal U}^{IJ}=\frac{1}{g^2}\int d^4 x Z^{(I)\,a}_MZ_M^{(J)\,a}\,,
\end{equation}
and performing the integration over the zero modes in (\ref{partition function of the bion}) results in
\begin{eqnarray}
\nonumber
Z_{\mbox{\scriptsize bion}}&=& \int \prod_{I=1}d\gamma_I\left(\mbox{Det}\, {\cal U}\right)^{1/2}\exp[-{8\pi^2\over g^2}(1 + c g)]\exp[-S_{int}]\mbox{Det}'\left(M_{MN}\right)^{-1/2}\\
&&\times\mbox{Det}\left(M^{gh}\right)\prod_f^{n_f}\mbox{Det}\left(i\bar\sigma^{M}D_M^{\mbox{\scriptsize cl}}\right)\,,
\end{eqnarray}
and $\mbox{Det}'$ stands for the determinant with the zero modes removed; note also that $M^{gh}$ is a positive definite operator that has no zero modes, see, e.g., \cite{Vandoren:2008xg}. 

In the confining phase, the dual photon mass depends explicitly on  the coupling $g$. Therefore, we will be interested only in the parametric dependence of $Z_{\mbox{\scriptsize bion}}$ on $g$, and will not try to keep all ${\cal O}(1)$ numbers in $Z_{\mbox{\scriptsize bion}}$.\footnote{In any case, this would be a somewhat daunting task: the background of a BPS and anti-KK monopole is neither (anti-)self dual nor does it enjoy any enhanced  symmetry.} Simple dimensional analysis reveals that at large distances ${\cal U}^{IJ}$ takes the form
\begin{equation}
{\cal U}^{IJ}\sim\frac{1}{g^2}\left(C_{IJ}+D_{IJ}\frac{L}{|\vec x_{\mbox{\scriptsize BPS}}-\vec x_{\overline{\mbox{\scriptsize KK}}}|} + \ldots \right)\,,
\end{equation}
where $C_{IJ}$ and $D_{IJ}$ are dimensionless constants ${\cal O}(1)$. Since we are working in the small-$\S^1$ dilute gas regime, where $L\ll |\vec x_{\mbox{\scriptsize BPS}}-\vec x_{\overline{\mbox{\scriptsize KK}}}|$ (the bion constituents   are generally far apart, separated by a distance of order $L/g^2$, as we will see in what follows), we can neglect the second term above. This leaves us with a contribution of a factor $1/g$ to $Z_{\mbox{\scriptsize bion}}$ for each collective coordinate. We have $8$ collective coordinates and in total we find $\left(\mbox{Det}\, {\cal U}\right)^{1/2}\sim 1/g^8$. 

The determinants for gauge fields, fermions and ghosts are ultraviolet divergent. In order to regularize these determinants we use the Pauli-Villars regularization scheme. Simply, we divide the determinant of each operator $\mbox{Det} O$ by $\mbox{Det}( O+\mu^d)$, where $\mu$ is a cut-off mass and $d$ is the mass dimension of the operator. Since we have $8$ zero modes in the gauge sector, we obtain a factor of $\mu^8$ in the numerator.
Collecting everything,  we find
\begin{eqnarray}
\nonumber
Z_{\mbox{\scriptsize bion}}&\sim&\frac{\mu^8}{g^8} \int d^3 x_{\mbox{\scriptsize BPS}}d^3 x_{\overline{\mbox{\scriptsize KK}}}\left(L\, d \theta_{\mbox{\scriptsize BPS}}\right)\left(L\,d\theta_{\overline{\mbox{\scriptsize KK}}}\right) \exp[-{8\pi^2\over g^2}(1 + c g)]\exp[-S_{int}]\\
\label{semi-final integration to be done}
&&\times\mbox{Det}'_{\mbox{\scriptsize reg}}\left(M_{MN}\right)^{-1/2}\mbox{Det}_{\mbox{\scriptsize reg}}\left(M^{gh}\right)\prod_f^{n_f}\mbox{Det}_{\mbox{\scriptsize reg}}\left(i\bar\sigma^{M}D_M\right)_{\mbox{\scriptsize reg}}\,,
\end{eqnarray}
where $\sim$ denotes factors of ${\cal O}(1)$ and all the determinants in the above expression are regularized by the background Pauli-Villars method. Finally, we are ready to perform the integrations in (\ref{semi-final integration to be done})  once we determine the functional dependence of $S_{int}$ and $\prod_f^{n_f}\mbox{Det}\left(i\bar\sigma^{M}D_M\right)$ on $x_{\mbox{\scriptsize BPS}}$ and $x_{\overline{\mbox{\scriptsize KK}}}$.

\subsubsection{Computation of $S_{int}$}
\label{computationofsint}

This computation of the ``interaction energy" is a straightforward one. One simply plugs in the expressions of the BPS and $\overline{\mbox{KK}}$ magnetic field strength,   given by (\ref{magnetic field for the BPS monopole}) and  (\ref{explicit time dependence in kk monopole}), in the expression for $S_{int}$ from eqn.~(\ref{interaction}):
\begin{equation}
S_{int}=\frac{L}{g^2}\int d^3 x \vec B^{\overline{\mbox{\scriptsize KK}}}\left(\vec x-\vec x_{\overline{\mbox{\scriptsize KK}}}\right)\cdot \vec B^{\mbox{\scriptsize BPS}}\left(\vec x-\vec x_{\mbox{\scriptsize BPS}}\right)\,.
\end{equation}
Since the non-radial components of the magnetic field die away outside the monopole cores, it is enough to keep the radial components. However, one should keep the full expression of $F_2(r)$ in  (\ref{magnetic field for the BPS monopole}) and  (\ref{explicit time dependence in kk monopole}) if we want to avoid singularities at the locations of the monopoles. Using the change of variables $\vec x-\vec x_{\mbox{\scriptsize BPS}}=\vec y$, going to the spherical coordinates, and performing the integrations in the $\theta$ and $\phi$ coordinates we obtain:
\begin{eqnarray}
S_{int}=\frac{2\pi L}{g^2r}{\cal I}\left({r\over L}\right)\,,
\end{eqnarray}
where $r=|\vec x_{\mbox{\scriptsize BPS}}-\vec x_{\overline{\mbox{\scriptsize KK}}}|$, and ${\cal I}(x)$ is given by:
\begin{eqnarray}
\nonumber
&&{\cal I}(x)=\int_0^\infty dR \left[1-\left(\frac{xR}{\sinh(xR)}\right)^2 \right]\\
\nonumber
&&\times\left\{\frac{R^2-1}{2R^2}\left(\frac{1}{|R-1|}-\frac{1}{R+1}+x\coth(x(R+1))-x\coth(x|R-1|)\right) \right.\\
\nonumber
&&\left.+\frac{R+1-|R-1|}{2R^2} -\frac{1}{2R^2}\left[4xR+2(R+1)\log\left(1-e^{-2x(R+1)}  \right) \right.\right.\\
\nonumber
&&\left.\left.  -2|R-1|\log\left(1-e^{-2x|R-1|}  \right)-x(R+1)^2\coth(x(R+1))+x(R-1)^2\coth(x|R-1|)  \right. \right.\\
&&\left. \left.-\frac{\mbox{Li}_2\left(e^{-2x(R+1)}\right)}{x}+\frac{\mbox{Li}_2\left(e^{-2x|R-1|}\right)}{x} \right] \right\}\,,  
\end{eqnarray}
where $\mbox{Li}_n(x)$ is defined by the infinite sum $\mbox{Li}_n(x)=\sum_{k=1}^\infty\frac{x^k}{k^n}$. 
We can calculate ${\cal I}$ numerically to find that ${\cal I} \rightarrow 2$ as $r\gg L$, as expected. Hence, we finally have:
\begin{eqnarray}\label{final expression for Sint}
S_{int}\left(\vec x_{\mbox{\scriptsize BPS}}-\vec x_{\overline{\mbox{\scriptsize KK}}}\right)=\frac{4\pi L}{g^2|\vec x_{\mbox{\scriptsize BPS}}-\vec x_{\overline{\mbox{\scriptsize KK}}}|}\,.
\end{eqnarray}

\smallskip

\subsubsection{Computation of $\prod_f^{n_f}\mbox{Det}\left(i\bar\sigma^{M}D_M^{\mbox{\scriptsize cl}}\right)$}
\label{computationofdet}

In general, computing the fermionic determinant is one of the formidable tasks in field theory. However, here we use an approximation method based on the existence of fermionic zero modes in the background of instantons with nontrivial holonomy. This method was implemented  before to study the interaction between instantons and anti-instantons in the presence of light Dirac fermions \cite{Diakonov:1984vw,Velkovsky:1997fe}. The Dirac operator has an exact zero mode in the background of (anti)instanton. Then, the basic idea in dealing with the fermionic interactions is that the Dirac spectrum in the background of an instanton-anti-instanton field can be split into quasizero modes, which are linear superposition of the zero modes of the individual instantons, and non-zero modes.  

Before showing how this procedure works in our case, we first review the construction of adjoint zero modes in the background of instantons (calorons) with non trivial holonomy  defined on $\R^3\times \S^1$ \cite{GarciaPerez:2006rt} (also see \cite{Davies:1999uw} for an earlier consideration).
The fermionic adjoint zero modes $\Psi^{a}$ are solutions of the Euclidean $4D$ massless Dirac equation in the adjoint representation of $SU(2)$
\begin{equation}
D_{M}^{\mbox{\scriptsize cl}}\gamma^M \Psi=0\,,
\end{equation}
where $D_{M}^{\mbox{\scriptsize cl}}$ denotes the covariant derivative in the background of the caloron. In the Weyl  representation, the left and right handed modes reduce to the two-component spinors $\Psi_{\pm}$ satisfying
\begin{eqnarray}
\sigma_{ M}D_M^{\mbox{\scriptsize cl}}\Psi_-=0\, \quad \bar\sigma_{M}D_M^{\mbox{\scriptsize cl}}\Psi_+=0\,.
\end{eqnarray}
For self-dual solutions one finds $D_M^{\mbox{\scriptsize cl}}\Psi_-=0$. Then, one can easily show that the density $|\psi_-(x)|^2$ is a constant, and for non-compact spaces these are non-normalizable solutions. To construct the fermionic zero modes, one uses the map between adjoint zero modes and (anti)self-dual deformations of a self-dual background (for a review, see also \cite{Weinberg:2006rq}). Given such a deformation $\delta A_M$, one can first impose the background gauge condition $D_M^{\mbox{\scriptsize cl}}\delta A_M=0$, and then one can show that
\begin{equation}
\Psi_+=\delta A_M \sigma_M {\cal V}
\end{equation}
is a zero mode for any constant 2-spinor ${\cal V}$. Using the ADHM construction, the authors of \cite{GarciaPerez:2006rt} were able to show  that the zero-mode $\Psi_+$ can be constructed by linear combinations of contributions adding up to the total caloron field strength. As was pointed above, large calorons can be broken into their  fundamental constituent monopoles. In this case, it was shown in \cite{GarciaPerez:2006rt} that $\Psi_+$, close to the center of the constituent monopole, is given by:
\begin{equation}\label{zero mode for monopoles}
\Psi^a_+(\vec x-\vec x_{\mbox{\scriptsize mon}})=\sigma_i B^a_i\left(\vec x-\vec x_{\mbox{\scriptsize mon}}\right){\cal V}\,,
\end{equation}
with $a$ is the color index, $\vec x_{\mbox{\scriptsize mon}}$ is the position of the corresponding monopole, $B_i^a$ is its magnetic field, and $\Psi_+$ is normalized as $\int d^4 x |\Psi_+(x)|^2=1$. Moreover, because of the Euclidean CP invariance we find that $\Psi_+^c\equiv -i\tau_2\psi_+^\dagger$ serves as a second independent zero-mode solution. The final picture consists of well-separated monopole constituents of the caloron, with two fermionic zero-modes per Weyl  fermion, given by (\ref{zero mode for monopoles}), carried by each constituent monopole.\footnote{Even more directly, without considering calorons, one  notes  that eqn.~(\ref{zero mode for monopoles}) gives the supersymmetric zero modes of the BPS- and KK-monopole solutions. While our theory has no supersymmetry, the BPS-limit fermion zero modes can be obtained using supersymmetry as an auxiliary tool. We further note that the ``almost"-BPS nature of the BPS and KK monopole solutions relevant in our case does not affect the leading long-distance behavior of the fermion zero modes. First, the index theorem \cite{Nye:2000eg,Poppitz:2008hr} does not depend on the existence of a mass gap for $A_4$ (i.e., it does not require (anti-)self-duality of the background) and, second, the long-distance asymptotics of the fermion zero mode is  the one given by (\ref{zero mode for monopoles}), as can be seen from the expression for the general  non-BPS monopole adjoint zero modes \cite{Jackiw:1975fn}. }

In the basis spanned by the zero modes the fermionic determinant in the bion background   is given by
\begin{equation}
 \prod^{n_f}\mbox{Det}\left(i\bar\sigma^{M}D_M^{\mbox{\scriptsize cl}}\right)\cong|T|^{2n_f}\prod^{n_f}\mbox{Det}'\left(i\bar\sigma^{M}D_M^{\mbox{\scriptsize cl}}\right)
\end{equation}
with the matrix element:
\begin{equation}\label{matrix element for hopping}
T=\int d^4 x \bar\Psi_{\overline{\mbox{\scriptsize KK}}}\,\bar \sigma_M D_M \Psi_{\mbox{\scriptsize BPS}}\,,
\end{equation}
and $\mbox{Det}'$ denotes the non-zero modes contribution to the determinant.  The matrix element has the meaning of a hopping amplitude for fermions between the BPS and $\overline{\mbox{KK}}$ monopoles that constitute the bion. Using the dotted notation, the expression (\ref{matrix element for hopping}) can be written explicitly as
\begin{equation}
T=\int d^4 x \bar \Psi_{\dot \alpha}^{\overline{\mbox{\scriptsize KK}}\,b}\left(\vec x-\vec x_{\overline{\mbox{\scriptsize KK}}}\right)\left[\bar \sigma_{M\,\dot \alpha\beta}\partial_M\Psi_{\beta}^{\mbox{\scriptsize BPS}\,b}\left(\vec x-\vec x_{\mbox{\scriptsize BPS}}\right)+\epsilon^{bcd}A_{M}^{\mbox{\scriptsize cl}\,c}\bar \sigma_{M\,\dot\alpha\beta}\Psi^{\mbox{\scriptsize BPS}\,d}_{\beta}\left(\vec x-\vec x_{\mbox{\scriptsize BPS}}\right) \right]\,
\end{equation}
where $\dot \alpha\,, \beta=1,2$ are the spinor indices. Since the color indices of the long range part of $\Psi$ are fixed in the $\tau_3$ direction, the second term in the above expression vanishes because of the antisymmetry of $\epsilon^{bcd}$. Hence, we find:
\begin{equation}
T=L\rho_{jik}\int d^3 x B_j^{\overline{\mbox{\scriptsize KK}}}\left(\vec x-\vec x_{\overline{\mbox{\scriptsize KK}}}\right)\partial_i B_k^{\mbox{\scriptsize BPS}}\left(\vec x-\vec x_{\mbox{\scriptsize BPS}}\right)\,,
\end{equation}
and $\rho_{jik}=\bar{\cal  V}^{\overline{\mbox{\scriptsize KK}}}\bar \sigma_j\bar\sigma_i\sigma_k {\cal V}^{\mbox{\scriptsize BPS}}$. Using the change of variables $\vec x-\vec x_{\overline{\mbox{\scriptsize KK}}}=\vec y$, and noting that $\partial B_k^{\mbox{\scriptsize BPS}}\left(\vec y+\vec r\, \right)/\partial y^i= \partial B_k^{\mbox{\scriptsize BPS}}\left(\vec y+\vec r \,\right)/\partial \vec r^{\,i}$, where $\vec r=\vec x_{\overline{\mbox{\scriptsize KK}}}-\vec x_{\mbox{\scriptsize BPS}}$, we obtain:
\begin{eqnarray}
\nonumber
T&=&L\rho_{jik}\frac{\partial}{\partial \vec r^{\,i}}\int d^3 y B_j^{\overline{\mbox{\scriptsize KK}}}(\vec y)B_k^{\mbox{\scriptsize BPS}}\left(\vec y+\vec r\,\right)\\
&=&L\rho_{jik}\frac{\partial}{\partial \vec r^{\,i}}\left(\frac{\delta_{jk}}{3}\frac{g^2}{L}S_{int}\right)=\frac{4\pi L }{3\, r^2 }\rho_{jij}\hat r^i\,,
\end{eqnarray}
with $\hat r^i\equiv \vec r^{\,i}/r$. Finally, we regularize this determinant by dividing by $\prod^{n_f}\mbox{Det}\left(i\bar\sigma^{M}(D_M^{\mbox{\scriptsize cl}}+\mu)\right)$ to obtain a factor of $1/\mu$ for each  {\em would be} zero mode. Hence, the regularized fermionic determinant in the background of a BPS-$\overline{\mbox{KK}}$ pair reads
\begin{equation}\label{final expression of fermionic determinant}
 \prod^{n_f}\mbox{Det}_{\mbox{\scriptsize reg}}\left(i\bar\sigma^{M}D_M^{\mbox{\scriptsize cl}}\right)\cong\left|\frac{4\pi L }{3\mu\, |\vec x_{\mbox{\scriptsize BPS}}-\vec x_{\overline{\mbox{\scriptsize KK}}}|^2 }\rho_{jij}\hat r^i\right|^{2n_f}\prod^{n_f}\mbox{Det}'_{\mbox{\scriptsize reg}}\left(i\bar\sigma^{M}D_M^{\mbox{\scriptsize cl}}\right)\,.
\end{equation}

\smallskip

\subsubsection{Computation of $Z_{\mbox{\scriptsize bion}}$}
\label{computationofzbion}

Now, we are in a position to put everything together. Plugging  (\ref{final expression for Sint}) and (\ref{final expression of fermionic determinant}) into (\ref{semi-final integration to be done}) we obtain for the bion partition function:
\begin{eqnarray}
Z_{\mbox{\scriptsize bion}} &\sim&\\
&&  \frac{I_B}{g^8}\left(\mu^4L\right)^2 \left(\frac{4\pi}{3\mu L}\right)^{2n_f}e^{-{8\pi^2\over g^2}(1 + c g)}\mbox{Det}_{\mbox{\scriptsize reg}}'\left(M_{MN}\right) \mbox{Det}_{\mbox{\scriptsize reg}}\left(M^{gh}\right) \prod^{n_f}\mbox{Det}_{\mbox{\scriptsize reg}}'\left(i\bar\sigma^{M}D_M^{\mbox{\scriptsize cl}}\right)\, . \nonumber
\label{final expression for Zbion}
\end{eqnarray}
By using the change of variables $\vec r=\vec x_{\overline{\mbox{\scriptsize KK}}}-\vec x_{\mbox{\scriptsize BPS}}$, and then going to  spherical coordinates,   the integral $I_B$ (defined below)  over the bion constituent collective coordinates becomes:
\begin{eqnarray}
\label{int}
I_B &\equiv& \int d^3x_{\mbox{\scriptsize BPS}} d^3x_{\overline{\mbox{\scriptsize KK}}} \left|\rho_{jij}\hat r^i\right|^{2n_f}
\exp\left[-4\frac{\pi L}{g^2 |\vec x_{\mbox{\scriptsize BPS}}-\vec x_{\overline{\mbox{\scriptsize KK}}}|}-4n_f\log\frac{|\vec x_{\mbox{\scriptsize BPS}}-\vec x_{\overline{\mbox{\scriptsize KK}}}|}{L}\right] \nonumber \\ &=& V{\cal A}\int _{r_{min}}^{\infty}dr r^2 \exp\left[-V_{\mbox{\scriptsize eff}}\right]\,,
\end{eqnarray}
with the effective potential:
\begin{equation}
V_{\mbox{\scriptsize eff}}=\frac{4\pi L}{g^2 r}+4n_f\log\left(\frac{r}{L}\right)\,,
\end{equation}
and:
\begin{equation}
{\cal A}=\int_{0}^{2\pi}\int_0^\pi d\theta d\phi \sin\theta \,  \left|\rho_{jij}\hat r^i(\theta,\phi)\right|^{2n_f}\,,
\end{equation}
while $V$ is the $3$-space volume\footnote{Recall that, as explained in Section \ref{bionplasma}, this factor should be dropped from $Z_{bion}$ when computing the dual photon mass (\ref{photon mass}), as it is already included in the summation over bions in the bion plasma partition function (\ref{zdual}).} and $r_{min} \sim 1/L$ denotes the fact that we are working in a dilute gas approximation and in the effective
long-distance field theory, hence the BPS and $\overline{\rm{KK}}$ monopoles can not be ``on top" of each other. 

The stability of the bion depends on the existence of a local minimum of the effective potential at $r_*$, provided that $r_*\gg L$, so that we can trust our long-distance effective theory. The stability radius is given by $r_*={\pi L \over g^2 n_f} $, hence our treatment is valid as long as $g \ll 1$. 
Next, the integral in the $r$ variable can be performed by going to dimensionless variables $x = r/r_*$, giving, up to numerical constants that we omit, and using $r_{min}/r_*\sim g^2 n_f$:
\begin{eqnarray}
\label{IB}
I_B &\sim&  {V L^3 \over g^{6 - 8 n_f}} \int\limits_{ g^2 n_f}^\infty d x \; x^{2 - 4 n_f} e^{- {4 n_f \over x}} \sim {V L^3 \over g^{6 - 8 n_f}}.
 \end{eqnarray}
In the last equality we used the fact that   the $x$-integral, for the  nonsupersymmetric and asymptotically free  cases of interest $n_f=2,3,4,5$, is saturated by values $x \sim 1$ and is independent
of its lower limit, since the integrand is suppressed for small $x$ (i.e. the BPS and ${\overline{{\rm KK}}}$ monopoles do not overlap).  It is clear from (\ref{IB}) that the ``size" of the bion molecule $r_* = {\pi L \over g^2 n_f}$ is larger than the Compton wavelength of the ``Higgs" field $A_4^3$, of order $L/g$, and we are justified in ignoring the ``electric" interactions between the bion constituents. The relevant scales for the bion molecule are depicted on Figure \ref{fig:bionsdraw}.

The  $g^2$ and  $n_f$ dependence of the integral over the quasizero mode  (\ref{IB}) can also be obtained\footnote{We thank M. \" Unsal for sharing his unpublished notes on this topic.} by considering the connected correlator of two monopole-instanton induced 't Hooft-vertex operators, as in \cite{Unsal:2007jx}. The reason the method used in  \cite{Unsal:2007jx} applies is that the bion binding dynamics is long-distance compared to the $1/L$ ``UV-cutoff" scale where the 't Hooft vertices (i.e., the monopole and KK monopole ``disorder" operators with fermions attached) are generated, hence, computing their connected correlators is equivalent to course-graining the UV lagrangian; we note that the effective field theory point of view will also be advocated below.

At this stage, one can compute the one-loop determinants, which get contributions only from non-zero modes, to find an additional logarithmic dependence on $\mu$ coming from the ultraviolet divergence. The ultraviolet divergent terms combine with the zero-mode terms to give the coefficient of the one-loop $\beta$ function, and the bare coupling $g$ and the Pauli-Villars regulator $\mu$ combine to give a running coupling constant $g(\mu)$. In fact, at two-loop order the non-zero mode determinants combine with the bare charge to give the two-loop $\beta$ function in the exponent  in (\ref{final expression for Zbion}), and the one-loop running coupling in the pre-exponent. While it is difficult to perform such a calculation explicitly, one can argue, using the renormalizability of the theory and the fact that $1/L$ is the only mass scale ever introduced, that all couplings should be taken to be running couplings at the scale $1/L$. 

An additional argument, based on effective field theory, is appropriate since we are working at small $L$. The BPS and leading twisted (KK) monopole solutions that constitute the bions of smallest charge and action, i.e. the one most relevant at small $L$, involve only the lowest Kaluza-Klein modes (with KK numbers $0, \pm 1$, see the explicit solutions (\ref{explicit time dependence in kk monopole})) of the fields and can be effectively described by a 3D theory that  only involves these lowest modes.\footnote{Such a theory would be relatively straightforward to obtain via ``deconstruction"---see  \cite{Poppitz:2003uz}  for a construction of the tower of twisted monopole solutions in such a framework. Since deconstruction approximates the ``extra" dimension only by a finite number of Kaluza-Klein modes, the corresponding tower of twisted monopoles is also finite.}
The coupling in this 3D theory is given by $g(L)/L$ (we do not distinguish between the energy scales $\pi/L$ or $2 \pi/L$  here, a difference that will only introduce an inessential correction). Since the 3D theory is Higgsed at the scale $\pi/L$, there is no further running of the 3D coupling and all the physics should be expressed in terms of $g(L)$,  obeying the usual (unbroken) 4D renormalization group equation.

Thus, we argue that  the dependence of  $Z_{\scriptsize \mbox{bion}}$ on $g$  is given to two-loop order by:
\begin{equation}
\label{finalzbion}
Z_{\scriptsize \mbox{bion}} \left(g(L)\right)\sim \frac{1}{g_{\scriptsize\mbox{1-Loop}}^{14-8n_f}\left(L\right)} \; e^{-{8\pi^2\over g_{\scriptsize \mbox{2-Loop}}^2\left(L\right)} (1 + c g_{\scriptsize \mbox{2-Loop}} \left(L\right))}
\end{equation}  
where $\sim$  denotes coefficients that play no role in determining the dependence of the dual-photon mass ${\cal M}$ on the energy scale. 

\subsection{Dual photon mass and previous small-$L$  ``estimates" of conformal window}
\label{dualphotonmass}

In this subsection, we determine the dependence of the photon mass on the $S^1$ size to two-loop order. The dual photon mass is given by eqn.~(\ref{photon mass}), which after substituting (\ref{finalzbion}), reads:
\begin{eqnarray}
{\cal M}&=&4 \pi\sqrt{\frac{ 8 Z_{\mbox{\scriptsize bion}}(g)}{ g^2L^2}} \\
&\sim& \frac{1}{L}\exp\left[-\frac{4\pi^2}{g_{\scriptsize\mbox{2-Loop}}^2(L)}(1 + c g_{\scriptsize\mbox{2-Loop}}) +\left(2n_f-4\right)\log g_{\scriptsize\mbox{1-Loop}}^2(L)\right]~,\nonumber
\end{eqnarray}
and $g(L)$ is the running coupling at the energy scale $1/L$.
Plugging the appropriate loop order of (\ref{beta function to two loop order}) into ${\cal M}$ (recall that $\beta_0 = (22 - 4 n_f)/3, \beta_1 = (136-64 n_f)/3$), we obtain:
\begin{eqnarray}
\nonumber
\label{bionmassL}
{ {\cal{M}} \over \Lambda }  &\sim&\exp\left[
- {\beta_0\over 4}\left( \log {1\over \Lambda^2 L^2}\right) \left(1 - \frac{\beta_1}{\beta_0^2}\frac{\log\log {1\over\Lambda^2L^2} }{ \log {1\over \Lambda^2 L^2} }\right)^{-1} -\log  \Lambda L+\left(4- 2n_f\right)\log \log {1\over \Lambda^2 L^2}   \right] \times \nonumber \\
&& \times e^{ - 2 \pi c \sqrt{ \log \left(1\over \Lambda L\right)^{\beta_0 \over 2} (1 + \ldots)} } \nonumber \\
&\sim&(\Lambda L)^\frac{\beta_0 - 2}{2} e^{- 2 \pi c \left({\beta_0 \over 2} \log {1 \over \Lambda L}\right)^{1/2}} \left(\log {1 \over \Lambda L} \right)^ { 4 - 2n_f - {\beta_1 \over 4 \beta_0}}  ~,
\end{eqnarray}
where we left out further subleading, at small $\Lambda L$, contributions. 
Recalling that $\beta_0 = (22 - 4 n_f)/3$, we find:
\begin{equation}
\label{bionmassL2}
{ {\cal{M}} \over \Lambda}  \sim (\Lambda L)^\frac{8 - 2 n_f}{3} e^{ - 2 \pi \tilde{c} \left( \log {1 \over \Lambda L} \right)^{1/2} } \times ({\rm less \; relevant \; contributions})~,
\end{equation}
where we use the positive number $\tilde{c} = 2 \pi c \sqrt{\beta_0/2}$.

We note that in the limit of asymptotically small $L\ll 1/\Lambda$, where the perturbative calculation is justified, the  correction to the leading semiclassical result $\sim (\Lambda L)^{8 - 2 n_f\over 3}$ is dominated by dependence\footnote{While the analytic expansion of eqn.~(\ref{higgscorrection}) of the non-BPS action is only valid for asymptotically  small $g \sim m_H/m_W \ll 10^{-3}$, see \cite{Forgacs:2005vx}, the numerical results 
 for the $m_H/m_W\sim g$ dependence of the action show that at weak coupling, $g \le 1$, the action is a monotonically increasing and approximately linear function, hence our conclusion is valid throughout the weak-coupling regime.} of the bion action on the nonzero Higgs mass, $\sim e^{- 4 \pi^2 c/g(L)}$.  As the size $L$ is increased, $g(L)$ increases, hence the exponential decreases---and the corresponding ``Higgs contribution" to the dual photon mass increases. For $n_f < 4$ and $n_f> 4$ this effect does not change the leading behavior dictated by the first factor on the r.h.s. of (\ref{bionmassL2}). 
However, for the four Weyl adjoint theory,  $n_f =4$, where the leading dependence of $\cal{M}$ on the $S^1$ size vanishes, we find that the next leading contribution to ${\cal{M}}\over \Lambda$, shown in (\ref{bionmassL2}) is an increasing function of $L$. The other terms shown in (\ref{bionmassL}) and omitted in (\ref{bionmassL2}) do not change this conclusion; this is most easily seen from the fact that their dependence on the gauge coupling is power-law, rather than exponential.

Thus, the dual photon mass ${\cal M}(n_f=4)$ increases with increasing $L$. Since the bion plasma density is proportional to the square of the dual photon mass, this means that the topological excitations do not dilute away in the decompactification limit---at least for sufficiently small $\Lambda L$, where this calculation is valid. 
Thus, according to the conjecture of  \cite{Poppitz:2009uq}, which ties conformality on $\R^4$ to dilution vs. nondilution of the mass gap on $\R^3 \times \S^1$ at increasing $L$, QCD with $N=2$, and $n_f=4$ Weyl fermions in the adjoint representation should not exhibit  conformal behavior in the large $L$ limit. 
Taking the ``estimate" of   \cite{Poppitz:2009uq} at face value means that the conformal window should be $4 < n_f<11/2$, i.e., occur only for the $n_f=5$ Weyl adjoints theory.  There are loopholes in this argument, of course, pertaining to the approach to $\R^4$ and we will discuss them in the next section.

\section{Summary and discussion}
\label{discussion}
  
\begin{figure}[h]
 \begin{center}
\includegraphics[width=5in]{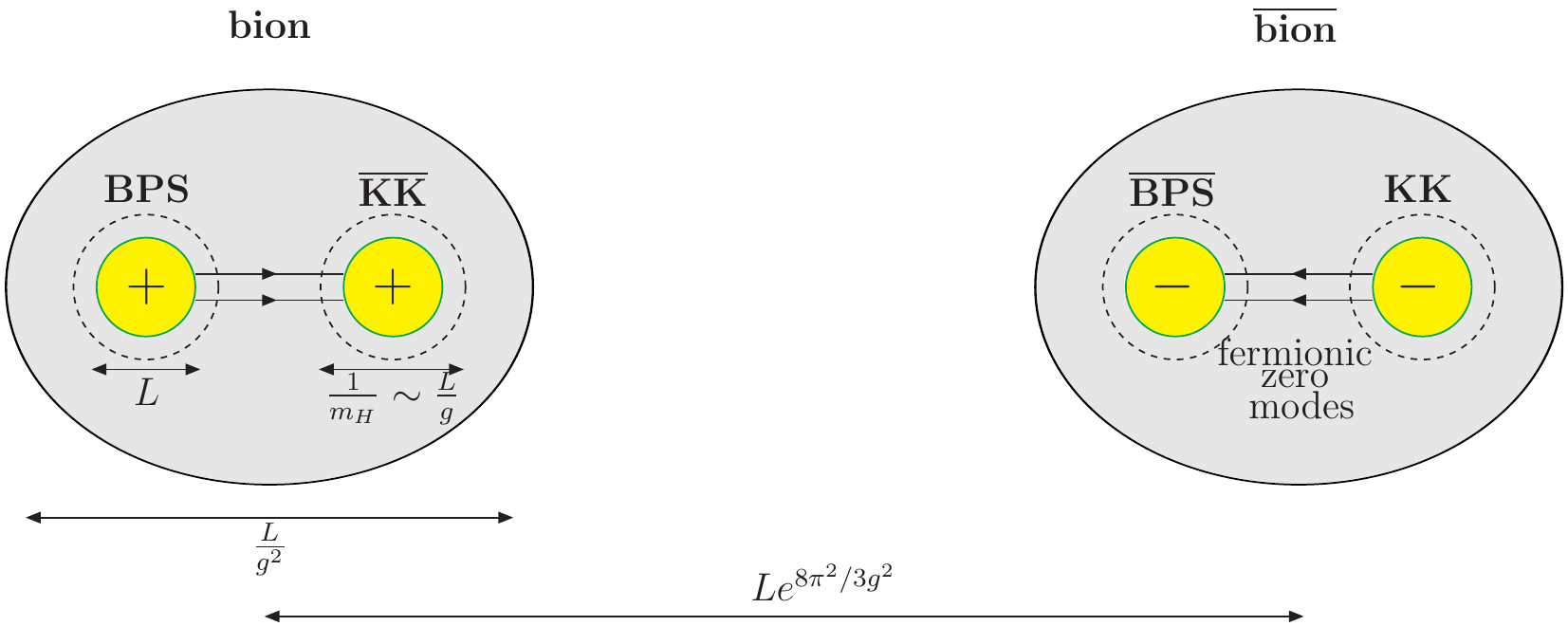}
\caption{The small-$L$, weak-coupling $g(L) \ll 1$,  hierarchy of scales in the bion plasma: the constituent monopole core size is $L$, the ``Higgs cloud" spreads over $L/g$, the bion size is $L/g^2$, and the typical distance between bions is $L e^{8 \pi^2 \over 3g^2}$. }
  \label {fig:bionsdraw}
 \end{center}
 \end{figure}

  In this paper, we studied in some detail 
 the $SU(2)$ gauge theory with $n_f$ massless adjoint Weyl fermions   on $\R^3 \times \S^1$, our main focus being 
  the bion mechanism of confinement of \cite{Unsal:2007jx}. We described  in detail the tools and approximations involved and discussed the stability of magnetic bions. The relevant scales in the problem at $L \ll \Lambda$ are  shown on  Figure~\ref{fig:bionsdraw}. We used methods and approximations familiar from QCD instanton calculations. 
 We also studied 
the behavior of  the mass gap (or string tension) as a function of $L$ at fixed $\Lambda$ for $n_f = 5,4,3,2$.
Already the earlier leading-order semiclassical result  \cite{Poppitz:2009uq}
 indicated  that  the $n_f = 5$ theory is perhaps conformal on $\R^4$, with (likely) a weakly-coupled infrared fixed point. The scenario,  at any finite $L$,  would then be that abelianization and abelian confinement take place, albeit with an exponentially small mass gap, $\sim {1 \over L}  e^{- {{\cal{O}}(1) \over g_*^2}}$, where $g_*$ is the small fixed point coupling. Thus, all mass scales in the theory approach zero as $L \rightarrow \infty$. 

 \begin{figure}[h]
 \begin{center}
\includegraphics[angle=0, width=2.5in]{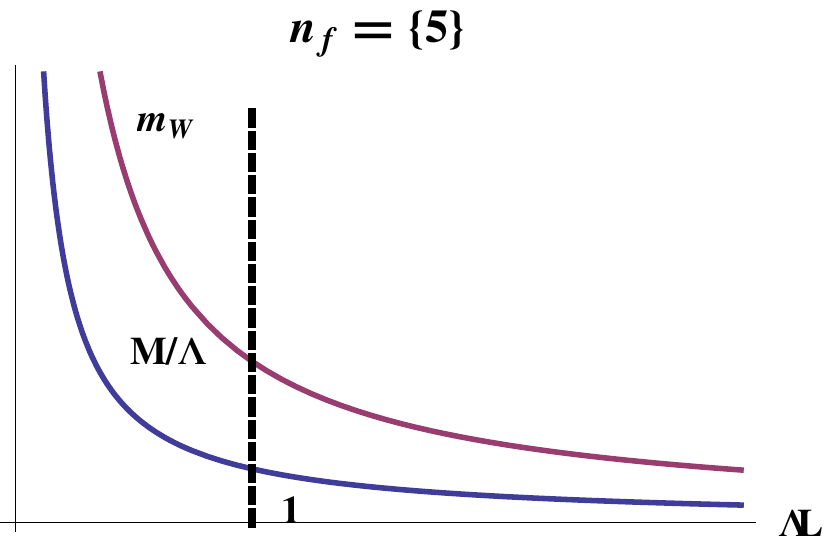}
\caption{Behavior of dual photon mass $M$ and W-boson mass ($\sim 1/L$) with $L$ for $n_f =5$. The behavior to the right of the dotted line at $\Lambda L \sim 1$, is based on the assumed existence of a weakly coupled infrared fixed point. This theory is thus expected to exhibit abelian confinement at any finite $L$ with an exponentially small string tension vanishing in the $\R^4$ limit.
 }
  \label {fig:nf5}
 \end{center}
 \end{figure}
   \begin{figure}[h]
 \begin{center}
\includegraphics[angle=0, width=2.5in]{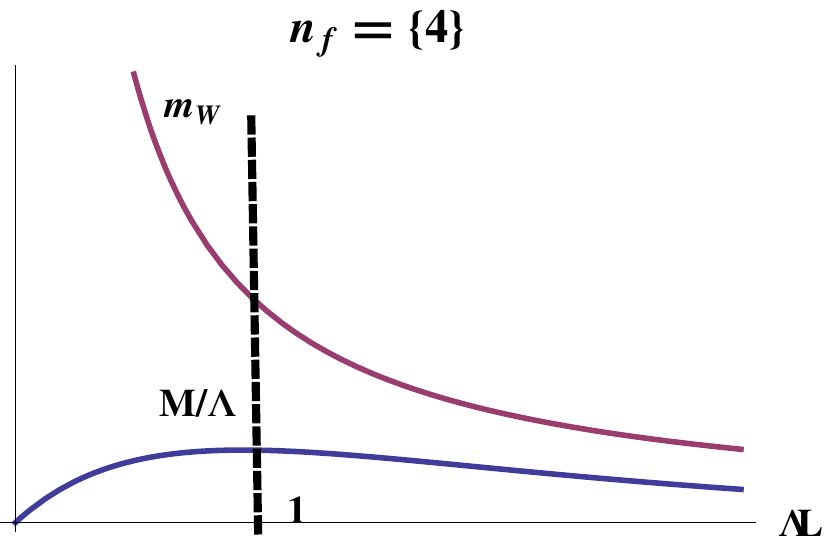}
\caption{Behavior of dual photon mass $M$ and W-mass with $L$ for $n_f =4$. The behavior to the right of the dotted line at $\Lambda L \sim 1$ is based on lattice evidence for the existence of a  weakly coupled infrared fixed point. If that is the case, the regime of semiclassical abelian confinement is expected to persist at any finite $L$, with an exponentially small string tension vanishing as $L \rightarrow \infty$.
 }
  \label {fig:nf4}
 \end{center}
 \end{figure}
  \begin{figure}[h]
 \begin{center}
\includegraphics[angle=0, width=2.5in]{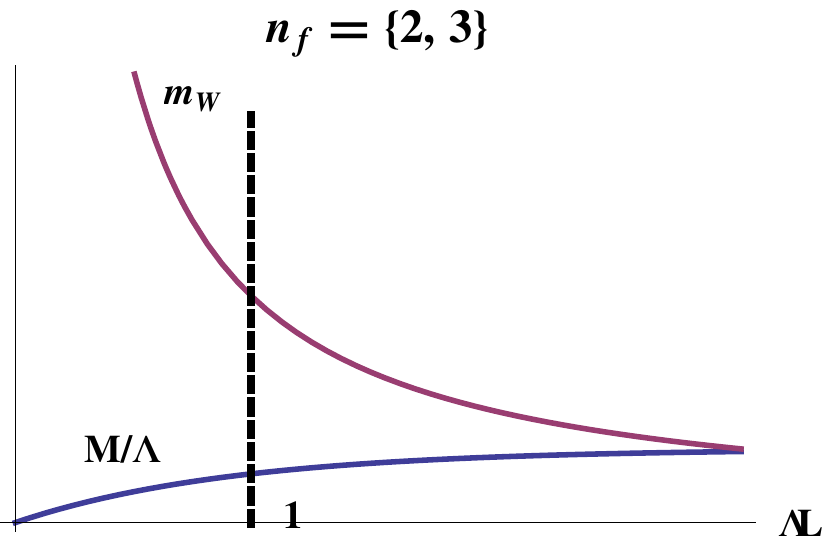}
\caption{Behavior of dual photon mass $M$ and W-mass with $L$ for $n_f =2,3$. The behavior to the right of the dotted line at $\Lambda L \sim 1$ assumes that a regime of nonabelian confinement sets in. The precise behavior at large $L$ is not known. The convergence to a common value of order $\Lambda$ is purely conjectural and is drawn similar to  the behavior of 
 quantities, analogous to our $m_W$ and $M$, studied in pure Yang-Mills theory on $T^3$ of size $L$ (see \cite{GarciaPerez:1993jw}, where, as $L$ is changed from $L \Lambda \ll 1$,  analytic methods were used, while for $L\Lambda \gg 1$ numerical studies were needed). 
 }
  \label {fig:nf23}
 \end{center}
 \end{figure}
 
The new result that we found here concerns the $n_f = 4$ theory, where the leading semiclassical result for the mass gap is  $L$ independent. The next-to-leading small-$\Lambda L$ behavior of the mass gap is  that it increases with $L$ at fixed $\Lambda$. Recent lattice studies, see 
\cite{recentlattice}, indicate that the $n_f=4$ theory is conformal on $\R^4$, apparently with a small fixed-point coupling and an anomalous dimension of the fermion bilinear at the fixed point in good agreement with one-loop perturbation theory. 
These results, combined with our analysis, then advocate for the following behavior of the $n_f=4$ theory on $\R^3 \times \S^1$.
As $L$ is increased, for $L \ll \Lambda^{-1}$, the coupling and the mass gap increase. As $L$  further increases past $L \sim \Lambda$, the coupling approaches the (weakly-coupled) fixed point $g_*$ and the mass gap  vanishes as ${1 \over L}  e^{- {{\cal{O}}(1) \over g_*^2}}$ as $L \rightarrow \infty$.\footnote{We note that this behavior is similar to the one expected  \cite{Poppitz:2009tw} for an $SU(2)$ gauge theory with a single four-index symmetric tensor ($j=2$) Weyl fermion---most likely a ``one-flavor" CFT on $\R^4$,  with a small, ``Banks-Zaks-ish," fixed-point coupling.} The perturbative and semiclassical analyses are then valid at any $L$, leading to the conclusion  that the theory  abelianizes and confines at any finite $\S^1$, albeit with an exponentially small string tension  $\gamma \sim{1\over L^2} e^{- {{\cal{O}}(1) \over g_*^2}}$.
Observing this behavior in either the $n_f=4$ Weyl adjoint  or  the $j=2$ Weyl fermion $SU(2)$ theories  on the lattice is probably prohibitively hard, as lattice studies \cite{Hart:1996ac}  of the Polyakov model have found, due to the small density of monopoles in the broken phase. 

Further decreasing the number of Weyl adjoint flavors, we found (here and earlier \cite{Poppitz:2009uq})  that for 
the theories with $n_f = 2,3$ the mass gap increases with $L$ and the theories 
are most likely confining on $\R^4$ (unless suprises lurk at large $L$ for $n_f=3$).  
The increase of the mass gap with $L$ at small $L\Lambda$ indicates that the topological excitations responsible for confinement become non-dilute.  Whether field configurations  with two units of magnetic charge are relevant for the dynamics of confinement at $L> \Lambda$ cannot be decided with the semiclassical methods of this paper. The general lesson learned from the small-$L$ study, however, is that the fermion backreaction can alter the nature of the topological excitations generating  confinement. To study this at large $L$, existing lattice studies of the role of topological defects in the confinement mechanism in pure Yang-Mills theory (see \cite{Greensite:2011zz} for a recent review) would have to be extended to theories with light  adjoint fermions. 
 
 \acknowledgments 
  We thank Mithat \" Unsal   for useful discussions. 
 This work was supported by the  National Science and Engineering Council of Canada (NSERC).

\end{document}